\documentclass[superscriptaddress,notitlepage,nofootinbib,groupedaddress]{revtex4-2}
\usepackage{natbib}
\usepackage{physics}
\usepackage{xcolor}
\usepackage{amsmath, amsfonts}
\usepackage{siunitx}
\usepackage{graphicx}
\usepackage[english]{babel}
\usepackage[utf8]{inputenc}
\usepackage[autostyle]{csquotes}
\MakeOuterQuote{"}
\usepackage[shortlabels]{enumitem}
\usepackage{dsfont}
\usepackage{comment}
\usepackage{verbatim}
\usepackage{xpatch}

\usepackage[hidelinks]{hyperref}	
\usepackage[capitalize]{cleveref}

\newcommand{\ie}{\textit{i.e.}}
\newcommand{\eg}{\textit{e.g.}}
\newcommand{\set}[1]{\{#1\}}

\begin{document}

	\title{Improved Finite-Key Security Analysis of Quantum Key Distribution Against Trojan-Horse Attacks}
    \author{Álvaro Navarrete}
    \affiliation{EI Telecomunicación, Dept. of Signal Theory and Communications, University of Vigo, E-36310 Vigo, Spain}
    \affiliation{atlanTTic Research Center, University of Vigo, E-36310 Vigo, Spain}
    \author{Marcos Curty}
    \affiliation{EI Telecomunicación, Dept. of Signal Theory and Communications, University of Vigo, E-36310 Vigo, Spain}
    \affiliation{atlanTTic Research Center, University of Vigo, E-36310 Vigo, Spain}

	\begin{abstract}
		Most security proofs of quantum key distribution (QKD) disregard the effect of information leakage from the users' devices, and, thus, do not protect against Trojan-horse attacks (THAs). In a THA, the eavesdropper injects strong light into the QKD apparatuses, and then analyzes the back-reflected light to learn information about their internal setting choices. Only a few recent works consider this security threat, but predict a rather poor performance of QKD unless the devices are strongly isolated from the channel. Here, we derive finite-key security bounds for decoy-state-based QKD schemes in the presence of THAs, which significantly outperform previous analyses. Our results constitute an important step forward to closing the existing gap between theory and practice in QKD.
	\end{abstract}

	\maketitle

	\section{Introduction}
	Quantum key distribution (QKD)~\cite{scarani,lo,portmann2021security,bennett} is arguably the most mature practical application of quantum information science, allowing to establish information-theoretic secure communications between two distant parties (commonly known as Alice and Bob) by combining the distribution of quantum systems to generate symmetric cryptographic keys with the well-known one-time-pad encryption scheme~\cite{vernam1926}. Unlike classical methods, whose security typically relies on computational assumptions, the security of QKD is only based on quantum information principles, and thus protects against any potential eavesdropper (Eve) with unlimited computational power.
	
	Nevertheless, there are still important challenges that need to be overcome to being able to deploy secure and practical QKD networks worldwide. In particular, it is critical to close the existing gap between the theoretical models used to prove the security of the protocols and their real-world implementations. Any deviation between the actual functioning of the devices employed by Alice and Bob and the physical model that characterizes their behavior might be exploited by Eve to compromise the security of QKD. Indeed, a typical assumption in most security proofs of QKD, including even those of device-independent QKD~\cite{mayers2004self,acin2007device,vazirani2014fully,arnon2018practical}, is that Alice and Bob's devices do not leak any unwanted information about their internal settings to the quantum channel. Unfortunately, however, this requirement is very hard to guarantee in practice. For instance, Eve could perform a so-called Trojan-horse Attack (THA)~\cite{gisin2,vakhitov} by injecting bright light into Alice's transmitter to create side channels that might leak sensitive information about the generated signals.
	Moreover, this information leakage is not the only vulnerability that this kind of invasive attacks can provoke. Indeed, it has been recently demonstrated for some commercial lasers that the injection of bright light into their cavities might increase the intensity of the transmitted pulses~\cite{huang2019laser}.
	
	There exist two main complementary approaches to re-establish the security of QKD in the presence of THAs. From the experimental side, one should implement methods to detect and monitor any potential side channel in real time, as well as to improve the isolation of the involved devices. On the theory side, one needs to relax the strict assumptions of most current security proofs to incorporate the effect of these potential side channels in the security analysis. Indeed, this is the approach that has been recently considered for instance  in~\cite{lucamarini2015Practical,tamaki2016decoy,wang2018finite,pereira2019quantum,pereira2019quantum}. In particular, in~\cite{tamaki2016decoy} the authors analyzed the asymptotic security of decoy-state QKD~\cite{hwang2003quantum,lo2005decoy,wang2005beating} in the presence of information leakage from Alice's intensity and bit/basis encoding setups. This work has been later on extended to the realistic finite-key regime by Wang \textit{et al.}~\cite{wang2018finite,wang2021measurement}. Unfortunately, the resulting secret-key rate is relatively poor and severely affected by both finite-key and side-channel effects, unless the devices are strongly isolated from the channel. Moreover, all these results do not take into account the fact that Eve's injected light might also vary the intensity of Alice's signals.
	
	In this work, we analyze the finite-key security of two well-known decoy-state-based QKD protocols in the presence of THAs. Specifically, we consider a decoy-state-based BB84~\cite{bennett,hwang2003quantum,lo2005decoy,wang2005beating} scheme, for which we notably improve the results reported previously in the scientific literature~\cite{wang2018finite}, with respect to the achievable key rate and distance. Besides, we consider a decoy-state-based loss-tolerant (LT)~\cite{tamaki2014loss} scheme, whose single-photon implementation is known to deliver the same asymptotic secret-key rate like that of the BB84 protocol in the absence of device imperfections. In order to derive both security analyses, we use two main ingredients. First, we take advantage of novel concentration inequalities for sums of dependent random variables~\cite{kato} to bound the finite-key deviations. And, second, we make use of the concept of reference states recently introduced in~\cite{pereira2020quantum,navarrete2021practical}. This allows us to incorporate any potential information leakage and intensity variation from Alice's devices into the analysis. The only requisite is to certify a single experimental parameter that encapsulates all the imperfections, and which is directly related to the isolation of the QKD devices. In doing so, we can roughly double the maximum achievable distance at which Alice and Bob can distill a secret key in various realistic scenarios when compared to previous approaches.

	\section{Transmitted states}
	Let us consider first the standard decoy-state BB84 protocol~\cite{hwang2003quantum,lo2005decoy,wang2005beating} with three intensity settings. In each round Alice prepares a BB84 state whose bit/basis encoding $a\in\{0_Z,1_Z,0_X,1_X\}$ is chosen with probability $p_{a}$, and whose intensity is set to $\mu\in\set{\mu_0,\mu_1,\mu_2}$ with probability $p_{\mu}$. That is, for concreteness here we consider that the settings $a$ and $\mu$ are selected independently, but the analysis below can be straightforwardly generalized to the case in which different intensity settings $\mu$ and probabilities $p_{\mu}$ are chosen for each basis ($Z$ and $X$)~\cite{zhou2016making}. Besides, we do not assume any specific encoding, being the analysis valid for all of them, \eg, polarization, phase, or time-bin encoding.
	
	In an entanglement-based view of the protocol, and in the absence of any device imperfection or attack, the state generated by Alice's source in any given round reads
	\begin{equation}
	    \ket{\Psi}_{ABC}=\sum_{a,\mu}\sqrt{p_{a}p_{\mu}}\ket{R_a,R_{\mu}}_{A}\ket{\phi_{a,\mu}}_{BC},
	\end{equation}
	where the states $\ket{R_a,R_{\mu}}_{A}\equiv\ket{R_a}_{A_1}\otimes\ket{R_{\mu}}_{A_2}$ form an orthonormal basis of Alice's register, with system $A=A_1A_2$, and
	\begin{equation}\label{eq:state_phi_a_mu}
	    \ket{\phi_{a,\mu}}_{BC}=\sum_{n=0}^{\infty}\sqrt{p_{n|\mu}}\ket{n}_{C}\ket{n_a}_{B}.
	\end{equation}
	The coefficients $p_{n|\mu}=e^{-\mu}\mu^{n}/n!$ denote the photon-number statistics corresponding to the intensity setting $\mu$, and $C$ is a purifying system not accessible to the parties such that $\Tr_{C}\set{\ket{\phi_{a,\mu}}_{BC}\bra{\phi_{a,\mu}}}=\sum_{n}p_{n|\mu}\ket{n_{a}}_{B}\bra{n_{a}}$, being $\ket{n_a}$ an $a$-encoded $n$-photon state. As standard, we consider that Alice selects her settings $a$ and $\mu$ with the pre-defined probabilities by performing projective measurement with elements $\{\ket{R_a,R_{\mu}}_{A}\}$ on her register. In particular, to simplify the notation we set $\ket{R_{0_Z}}=\ket{0}$, $\ket{R_{1_Z}}=\ket{1}$, $\ket{R_{0_X}}=\ket{2}$ and $\ket{R_{1_X}}=\ket{3}$.
	
	The absence of correlations between the generated states associated to different rounds implies that the global state of all the $N$ protocol rounds delivered by Alice's source factors as $\ket*{\Psi^{N}}_{ABC}=\bigotimes_{u=1}^{N}\ket{\Psi}_{ABC}^{u}$, where the round index $u$ refers to each system and state. In what follows, however, we will omit the index $u$ from the systems and states whenever it is clear that we refer to a particular round for simplicity of notation.
	
	\begin{figure}[ht]
		\centering
		\includegraphics[width=0.65\textwidth]{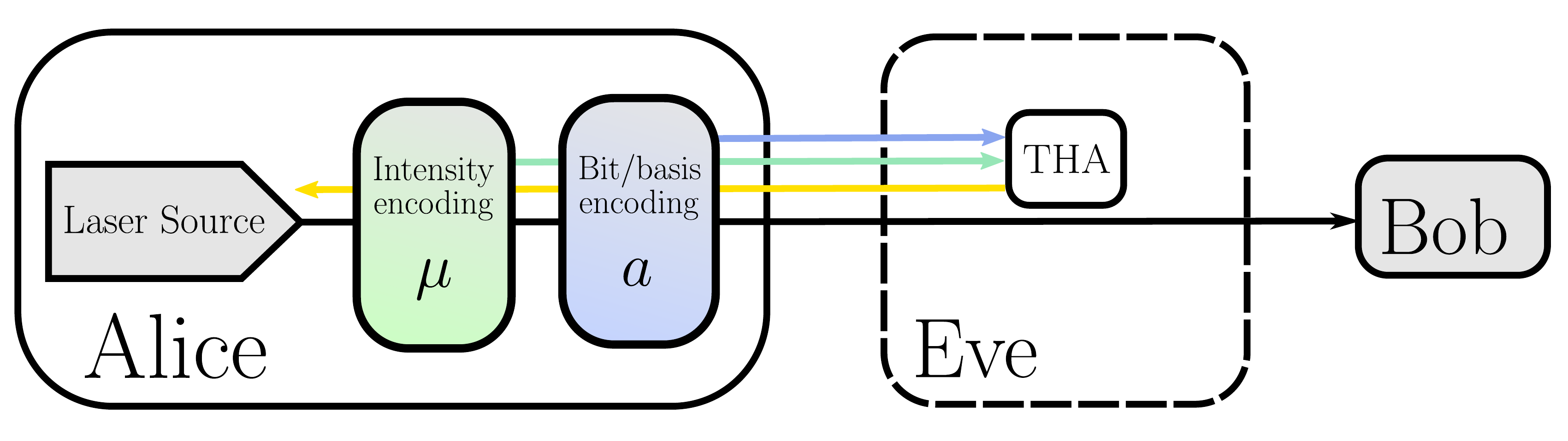}
		\caption{Schematic representation of a THA. Eve injects a photonic state (represented in the figure with a yellow arrow) into Alice's transmitter and analyzes the back-reflected light (represented in the figure with blue and green arrows), which might carry information about the internal configuration of Alice's bit/basis and intensity encoding setups. After that, Eve performs an arbitrary joint measurement on her own systems and Alice's transmitted pulses, and she decides the appropriate quantum states to be re-transmitted to Bob.} \label{fig:THA}
	\end{figure}
	Now let us consider that Eve injects an arbitrary photonic system into Alice's device with the aim of learning some information about both her bit/basis encoding ($a$) and intensity ($\mu$) choices by analyzing the back-reflected light, as illustrated in~\cref{fig:THA}. We have that the most general state describing all the quantum systems involved in this particular scenario after Eve's probe system interacts with Alice's bit/basis and intensity encoding setups can be written as
	\begin{equation}\label{eq:general_state_one_round}
	    \ket*{\Psi}_{ABCE}=\sum_{a,\mu}\sqrt{p_{a}p_{\mu}}\ket{R_a,R_{\mu}}_{A}\ket{\psi_{a,\mu}}_{BCE},
	\end{equation}
	where $\ket{\psi_{a,\mu}}_{BCE}$ is the state that is actually generated when Alice selects $a$ and $\mu$, being $E$ the optical mode of the back-reflected photonic light produced by the THA. Note that in general $E$ can also include any other systems at Eve's hands (as well as other modes inaccesible to Eve). Precisely, here we consider the case in which this state can be written as
	\begin{equation}\label{eq:actual_virtual_state_amu}
		\ket{\psi_{a,\mu}}_{BCE}=\sum_{n=0}^{\infty}\sqrt{\tilde{p}_{n|\mu}}\ket{n}_{C}\ket{\tilde{n}_{a,\mu}}_{BE},
	\end{equation}
	where the exact form of $\ket{\tilde{n}_{a,\mu}}_{BE}$ depends on Eve's THA and is typically unknown, and so the probabilities $\tilde{p}_{n|\mu}$, which do not necessarily need to follow a Poissonian distribution. This might happen because, as already mentioned, apart from the leak of information about Alice's settings via the state $\ket{\tilde{n}_{a,\mu}}_{BE}$, Eve's attack might modify the behavior of Alice's laser~\cite{huang2019laser}.

	\section{Security proof}\label{sec:SolRT}
	Here we prove the security of the standard decoy-state BB84 protocol when the emitted states are given by~\cref{eq:general_state_one_round,eq:actual_virtual_state_amu}. For this, we shall assume that Alice only sends a signal to Bob once he has detected the signal corresponding to the previous round. In doing so, we guarantee that Eve's actions in the $u$-th round cannot be influenced by the variables $a_{u'}$ and $\mu_{u'}$ for any $u'>u$.
	
	Our starting point is the conditional probability that Alice selects a particular intensity $\mu$, the $Z$ basis, and there is a \textit{click} at Bob's side in the $u$-th round given all the classical information publicly announced by them up to that round. This quantity can be written as
	\begin{equation}\label{eq:gain_actual}
	\begin{split}
	    Q_{Z,\mu}^u&=\bra*{\Psi}\Pi_{Z}\otimes\Pi_{\mu}\otimes\hat{\mathcal{D}}_{\text{click}}^{u}\ket*{\Psi}_{ABCE},
	\end{split}
	\end{equation}
	where $\Pi_{Z}=\ketbra{0}_{A_1}+\ketbra{1}_{A_1}$, $\Pi_{\mu}=\ketbra{R_{\mu}}_{A_2}$, $\hat{\mathcal{D}}_{\text{click}}^{u}$ is Bob-Eve's measurement operator associated to observing a \textit{click} in the $u$-th round, which acts on systems $B$ and $E$ and depends on all the classical information publicly announced by Alice and Bob up to that round, and $\ket*{\Psi}_{ABCE}$ is given in~\cref{eq:general_state_one_round}.
	
	Importantly, we note that the standard decoy-state technique cannot be applied directly to this scenario to relate the probabilities $Q_{Z,\mu}^u$ to the single-photon yields for two main reasons. First, the side channel provokes that the $n$-photon yields could now depend on the intensity setting $\mu$, and, second, the statistics $\tilde{p}_{n|\mu}$ might be in general unknown, as already mentioned. 
	
	To overcome these two problems and be able to use the decoy-state technique, we define a virtual reference state~\cite{pereira2020quantum} for that round as
	\begin{equation}\label{eq:reference_state_1}
	    \ket{\Phi}_{ABCE}=\sum_{a,\mu}\sqrt{p_{a}p_{\mu}}\ket{R_{a},R_{\mu}}_{A}\ket{\phi_{a,\mu}}_{BCE},
	\end{equation}
	where, in this case, we can decide a convenient form for $\ket{\phi_{a,\mu}}_{BCE}$, which is
	\begin{equation}\label{eq:purified_reference_state}
	    \ket{\phi_{a,\mu}}_{BCE}=\sum_{n=0}^{\infty}\sqrt{p_{n|\mu}}\ket{n}_{C}\ket{n_{a}}_{BE}=\sum_{n=0}^{\infty}\sqrt{p_{n|\mu}}\ket{n}_{C}\ket{n_{a}}_{B}\ket*{\tau}_E.
	\end{equation}
	That is, in \cref{eq:purified_reference_state} the states $\ket{n_{a}}_{BE}$ consist in a part $\ket{n_{a}}_{B}$ that is perfectly characterized ---they are the states ideally defined in the protocol, see~\cref{eq:state_phi_a_mu}---, and a part $\ket*{\tau}_E$ that could be any state of system $E$ which does not depend on Alice's settings. In short, besides having no information leakage, the reference state given in~\cref{eq:reference_state_1} represents a perfect phase-randomized weak coherent pulse when tracing out systems $C$ and $E$. For this reference state, the probability that Alice selects the intensity $\mu$, the $Z$ basis, and there is a \textit{click} at Bob's side in the $u$-th round conditioned on the previous public announcements made by Alice and Bob is defined analogously to \cref{eq:gain_actual}, \ie,
	\begin{equation}
		Q^{u,{\rm ref}}_{Z,\mu}=\bra*{\Phi}\Pi_{Z}\otimes\Pi_{\mu}\otimes\hat{\mathcal{D}}_{\text{click}}^{u}\ket*{\Phi}_{ABCE}
	\end{equation}
	These reference gains $Q^{u,{\rm ref}}_{Z,\mu}$, for the different intensity settings, can be straightforwardly related to the probability of observing a single-photon \textit{click} in the $u$-th round in the reference scenario, namely the single-photon yield $Y_{1}^{u,{\rm ref}}$, by means of well-known analytical or numerical bounds~\cite{lo2005decoy,wang2005beating,lim2014concise,zhang2017improved}. Since the reference states are never sent in the actual implementation of the protocol, we cannot directly observe the quantities $Q^{u,{\rm ref}}_{Z,\mu}$. Fortunately, however, one can indirectly estimate them by using the following relation~\cite{pereira2020quantum,navarrete2021practical,zapatero2021security} 
	\begin{equation}\label{eq:RT_Ineq_Gain}
	    \sqrt{Q^{u,{\rm ref}}_{Z,\mu}Q_{Z,\mu}^u}+\sqrt{(1-Q^{u,{\rm ref}}_{Z,\mu})(1-Q_{Z,\mu}^u)}\geq \delta_u
	\end{equation}
	where $Q^{u}_{Z,\mu}$ is given in~\cref{eq:gain_actual}, and $\delta_u = \abs{\braket{\Psi}{\Phi}_{ABCE}}$, with
	\begin{equation}\label{eq:delta1}
	\begin{split}
	    \abs{\braket{\Psi}{\Phi}_{ABCE}} &= \abs{\sum_{a,\mu}p_{a}p_{\mu}\braket{\psi_{a,\mu}}{\phi_{a,\mu}}_{BCE}}\\
	    & = \abs{\sum_{a,\mu}p_{a}p_{\mu}\sum_n \sqrt{p_{n|\mu}\tilde{p}_{n|\mu}}\braket{\tilde{n}_{a,\mu}}{n_{a}}_{BE}}.
	\end{split}
	\end{equation}
	The inequality $\sqrt{p_1p_2}+\sqrt{(1-p_1)(1-p_2)}\geq \delta$, which following~\cite{zapatero2021security} we will call Cauchy-Schwarz (CS) inequality, leads to two different bounds that depend on $\delta$, namely $p_1\leq G^{\rm U}_{\delta}(p_2)$ and $p_1\geq G^{\rm L}_{\delta}(p_2)$, where
	\begin{align}
		G^{\rm U}_{\delta}(p)&=\begin{cases}
			g^{+}_{\delta}(p) \quad & p<\delta^2, \\
			1 \quad & \text{otherwise}, \\
		\end{cases} &
		G^{\rm L}_{\delta}(p)&=\begin{cases}
			g^{-}_{\delta}(p) \quad & p>1-\delta^2, \\
			0 \quad & \text{otherwise}, \\
		\end{cases}
	\end{align}
	and $g^{\pm}_{\delta}(p)=p+(1-\delta^2)(1-2p)\pm2\delta\sqrt{(1-\delta^2)p(1-p)}$.
	
	Now, similarly to the gains $Q_{Z,\mu}^u$ and $Q_{Z,\mu}^{u,{\rm ref}}$, we denote the conditional probability of a $n$-photon \textit{click} at Bob's side in the $Z$ basis given all the previous information announced by Alice and Bob up to the $u$-th round by
	\begin{align}
		Y_{n}^{u,{\rm ref}}&=\bra*{\Phi}\Pi_{Z}\otimes\Pi_{n_{C}}\otimes\hat{\mathcal{D}}_{\text{click}}^{u}\ket*{\Phi}_{ABCE} & \text{and} &&
		Y_{n}^{u}& =\bra*{\Psi}\Pi_{Z}\otimes\Pi_{n_{C}}\otimes\hat{\mathcal{D}}_{\text{click}}^{u}\ket*{\Psi}_{ABCE},
	\end{align}
	for the reference and actual scenarios, respectively, with $\Pi_{n_C}\equiv\ketbra{n}_{C}$. Note that here we are defining the yields as joint probabilities, \ie, they are not conditioned on sending a $n$-photon pulse. Focusing on the single-photon case, the yields $Y_{1}^{u,{\rm ref}}$ and $Y_{1}^u$ can be related, again, through the CS inequality, \ie, 
	\begin{equation}\label{eq:RT_Ineq_yields}
	    \sqrt{Y^{u,{\rm ref}}_{1}Y_{1}^u}+\sqrt{(1-Y^{u,{\rm ref}}_{1})(1-Y_{1}^u)}\geq \delta_u.
	\end{equation}

	Next we use the previous relations to prove the security of the protocol in the presence of a THA. In particular, we first estimate the number of successful rounds in which Alice transmitted a single-photon pulse. Then, we estimate the number of single-photon phase errors within Alice and Bob sifted key. Finally, based on these estimations, we calculate a lower bound on the length of the secret key.
    
	\subsection{Number of detected single-photon pulses}\label{subsec:single-photon-yield}
    Here we estimate the number of detected $Z$-basis rounds in which Alice transmitted a single-photon pulse, namely $M_{1}^{Z}$, from the observed number of $Z$-basis detections for the different intensity settings, namely $M_{\mu}^{Z}$. This can be done, as mentioned before, by applying the decoy-state idea combined with the reference technique. Below we describe briefly the process:
	\begin{enumerate}
	    \item \textit{Finite-key bounds:} We first use concentration inequalities for sums of dependent random variables to lower bound $M_{1}^Z$ from a sum of conditional probabilities $Y_1^u$ that runs on the rounds $u=1,\dots,N$. In particular, by applying say Kato's inequality~\cite{kato} one can lower-bound $M_{1}^Z$ as
		    \begin{equation}\label{eq:first-finite-key-bound}
		        M_{1}^Z\geq \bar{K}_{N,\epsilon}^{\rm L}\left(\sum_{u=1}^N Y^{u}_1\right),
		    \end{equation}
    	which holds except with probability $\epsilon$. The function $ \bar{K}_{N,\epsilon}^{\rm L}$ is defined in~\cref{appendix:ConcentrationBounds}.

	    \item \textit{CS inequality:} The lower bound in~\cref{eq:first-finite-key-bound} requires an estimation on the sum $\sum_{u=1}^{N} Y^{u}_1$. As mentioned at the beginning of this section, in the absence of a THA one could relate the single-photon yields $Y_1^u$ of each round directly to the gains $Q^{u}_{Z,\mu}$ through the decoy-state method. In the presence of a THA this is not possible, so we take advantage of the reference states. In particular, from \cref{eq:RT_Ineq_yields}, we know that each single-photon yield $Y^u_{1}$ in~\cref{eq:first-finite-key-bound} can be related to its corresponding reference yield $Y^{u,{\rm ref}}_{1}$ by means of the CS inequalty, \ie,
    	    \begin{equation}\label{eq:first-CS-bound}
    	        Y^u_{1}\geq G^{\rm L}_{\delta}\left(Y^{u,{\rm ref}}_{1}\right),
    	    \end{equation}
    	where the parameter $\delta$ represents any lower bound on the quantity $\delta_u$. Furthermore, due to the convexity of the function $G^{\rm L}_{\delta}(p)$, one can use Jensen's inequality~\cite{jensen1906fonctions} to obtain a lower bound on the sum of single-photon yields, \ie,
    	    \begin{equation}
    	        \sum_{u=1}^N Y^{u}_1 \geq N G^{\rm L}_{\delta}\left(\frac{1}{N}\sum_{u=1}^N Y^{u,{\rm ref}}_1\right),
    	    \end{equation}
    	which can be directly plugged into \cref{eq:first-finite-key-bound} to obtain
    	\begin{equation}\label{eq:Mz1L}
    		M_{1}^Z\geq \bar{K}_{N,\epsilon}^{\rm L}\left(N G^{\rm L}_{\delta}\left(\frac{1}{N}\sum_{u=1}^N Y^{u,{\rm ref}}_1\right)\right).
    	\end{equation}

	    \item \textit{Decoy-state technique:} For the reference states it is possible to write the single-photon yields as a linear combination of the different gains, \ie, we have that $Y^{u,{\rm ref}}_1 \geq F_{\rm D}(Q_{Z,\mu_1}^{u,{\rm ref}},\dots,Q_{Z,\mu_d}^{u,{\rm ref}})$, for a certain linear function $F_{\rm D}$ of the gains. Besides, due to the linearity of $F_{\rm D}$, we have that
	    \begin{equation}
	       \frac{1}{N}\sum_{u=1}^N Y^{u,{\rm ref}}_1 \geq F_{\rm D}\left(\frac{1}{N}\sum_{u=1}^NQ_{Z,\mu_1}^{u,{\rm ref}},\dots,\frac{1}{N}\sum_{u=1}^NQ_{Z,\mu_d}^{u,{\rm ref}}\right).
	    \end{equation}
	    For convenience, we write $\frac{1}{N}\sum_{u=1}^NY_{n}^{u,{\rm ref}}$ as
	    \begin{equation}\label{eq:Yield_redef}
			\frac{1}{N}\sum_{u=1}^NY_{n}^{u,{\rm ref}}=\frac{p_np_{Z}^Ap_{Z}^B}{N}\sum_{u=1}^N\tilde{Y}_{n}^{u,{\rm ref}},
	    \end{equation}
	    where $p_n=\sum_{\mu}p_{\mu}p_{n|\mu}$ and $\tilde{Y}_{n}^{u,{\rm ref}}$ is the conditional probability of observing a \textit{click} at Bob's side given that Alice's transmitted a $n$-photon pulse and both users selected the $Z$ basis in the reference scenario. Then, we have that $\frac{1}{N}\sum_{u=1}^N\tilde{Y}_{1}^{u,{\rm ref}}$ can be lower bounded by solving the following linear program (LP): 
	    \begin{equation}\label{eq:LP_Yields}
	    	\begin{split}
	    		\text{min} & \quad  \frac{1}{N}\sum_{u=1}^N\tilde{Y}_{1}^{u,{\rm ref}}\\
	    		\text{s.t.}  & \quad  \frac{1}{Np_{\mu}p_Z^Ap_Z^B}\sum_{u=1}^NQ_{Z,\mu}^{u,{\rm ref}} -\Lambda_\mu \leq \sum_{n=0}^{n_{\text{cut}}} p_{n|\mu} \left(\frac{1}{N}\sum_{u=1}^N\tilde{Y}_n^{u,{\rm ref}}\right) \leq \frac{1}{Np_{\mu}p_Z^Ap_Z^B}\sum_{u=1}^NQ_{Z,\mu}^{u,{\rm ref}}, \quad \forall \mu
	    	\end{split}
	    \end{equation}
	    where
	    \begin{equation}\label{eq:Lambda}
	        \sum_{n=n_{\text{cut}}+1}^{\infty} \frac{e^{-\mu}\mu^n}{n!}\left(\frac{1}{N}\sum_{u=1}^N\tilde{Y}_n^{u,{\rm ref}}\right) 
	        \leq 1-\sum_{n=0}^{n_{\text{cut}}} e^{-\mu}\frac{\mu^n}{n!}=:\Lambda_\mu,
	    \end{equation} 
	    \item \textit{CS inequality:} We can now bound the reference gains $Q_{Z,\mu}^{u,{\rm ref}}$ that are required to estimate the reference single-photon yields through the LP presented in the previous step by applying again the CS inequality, obtaining
	    \begin{equation}
	        \frac{1}{N}\sum_{u=1}^N G_{\delta}^{\rm L}(Q_{Z,\mu}^{u}) \leq \frac{1}{N}\sum_{u=1}^N Q_{Z,\mu}^{u,{\rm ref}} \leq \frac{1}{N}\sum_{u=1}^N G_{\delta}^{\rm U}(Q_{Z,\mu}^{u}).
	    \end{equation}
	    Besides, we can take advantage again of the convexity and concavity of the functions $G_{\delta}^{\rm L}$ and $G_{\delta}^{\rm U}$, respectively, to obtain
	    \begin{equation}\label{eq:Jensen2}
	       G_{\delta}^{\rm L}\left(\frac{1}{N}\sum_{u=1}^N Q_{Z,\mu}^{u}\right)  \leq \frac{1}{N}\sum_{u=1}^N Q_{Z,\mu}^{u,{\rm ref}} \leq G_{\delta}^{\rm U}\left(\frac{1}{N}\sum_{u=1}^N Q_{Z,\mu}^{u}\right).
	    \end{equation}

	    \item \textit{Finite-key bounds:} Finally we can apply again concentration inequalities to bound, with very high probability, the sum of gains from a function of the number of \textit{clicks} observed by Alice and Bob (see \cref{appendix:ConcentrationBounds}). That is,
	    \begin{equation}\label{eq:QfiniteBounds}
	       K_{N,\epsilon}^{\rm L}\left(M_{\mu}^{Z}\right) \leq \sum_{u=1}^N Q_{Z,\mu}^{u} \leq K_{N,\epsilon}^{\rm U}\left(M_{\mu}^{Z}\right),
	    \end{equation}
	    where $M_{\mu}^{Z}$ is the number of detections in which Alice selects the intensity $\mu$, and both she and Bob select the $Z$ basis. From~\cref{eq:Jensen2,eq:QfiniteBounds} we have that the constraints on the LP given by \cref{eq:LP_Yields} now take the form
	    \begin{equation}\label{eq:LP_Yields_Edited}
	    	\frac{1}{p_{\mu}p_Z^Ap_Z^B}G_{\delta}^{\rm L}\left(\frac{1}{N}K_{N,\epsilon}^{\rm L}\left(M_{\mu}^{Z}\right)\right) -\Lambda_\mu \leq \sum_{n=0}^{n_{\text{cut}}} p_{n|\mu} \left(\frac{1}{N}\sum_{u=1}^N\tilde{Y}_n^{u,{\rm ref}}\right) \leq \frac{1}{p_{\mu}p_Z^Ap_Z^B}G_{\delta}^{\rm U}\left(\frac{1}{N}K_{N,\epsilon}^{\rm U}\left(M_{\mu}^{Z}\right)\right).
	    \end{equation}

	\end{enumerate}
	By solving this LP, one can obtain a lower bound on $\frac{1}{N}\sum_{u=1}^N\tilde{Y}_{1}^{u,{\rm ref}}$. Finally, by combining \cref{eq:Yield_redef,eq:Mz1L}, one obtains a lower bound $M_{1}^{Z, {\rm L}}\leq M_{1}^Z$.
	
	In the following subsection we apply an analogous procedure to estimate the number of single-photon phase errors $M_{\rm ph,1}$. 
	
	\subsection{Number of single-photon phase errors}\label{subsec:single-photon-phase-error}
	We consider that Alice and Bob extract their secret keys from the detected rounds in which Alice prepares a single-photon pulse and both she and Bob select the $Z$ basis. We then define a virtual scenario in which Alice and Bob perform their measurements in the complementary basis for all of such key rounds, which means that Alice measures her system $A_1$ in the basis $\set{\ket{0_X},\ket{1_X},\ket{2},\ket{3}}$, where $\ket{j_X}=\frac{1}{\sqrt{2}}(\ket{0}+(-1)^j\ket{1})$.
	
	Let us rewrite the reference state given by \cref{eq:reference_state_1,eq:purified_reference_state} as
	\begin{equation}
		\ket{\Phi}_{ABCE}=\sum_{a,\mu}\sqrt{p_{a}p_{\mu}p_{n|\mu}}\ket{R_{a},R_{\mu}}_{A}\sum_{n\neq1}\ket{n}_{C}\ket{n_{a}}_{BE}  +\sum_{\mu}\sqrt{p_{\mu}p_{1|\mu}}\ket{R_{\mu}}_{A_2}\ket{1}_{C}\ket{\phi_{1}}_{A_1BE},
	\end{equation}
	with
	\begin{equation}\label{eq:reference_virtual2}
		\begin{split}
			\ket{\phi_{1}}_{A_1BE} = & \sqrt{p_{0_Z}}\ket{0}_{A_1}\ket{1_{0_Z}}_{BE} +\sqrt{p_{1_Z}}\ket{1}_{A_1}\ket{1_{1_Z}}_{BE}
			+\sqrt{p_{0_X}}\ket{2}_{A_1}\ket{1_{0_X}}_{BE}
			+\sqrt{p_{1_X}}\ket{3}_{A_1}\ket{1_{1_X}}_{BE}\\
			= & \sqrt{p_{0}^{\rm vir}}\ket{0_X}_{A_1}\ket{1_{0}^{\rm vir}}_{BE} 
			+\sqrt{p_{1}^{\rm vir}}\ket{1_X}_{A_1}\ket{1_{1}^{\rm vir}}_{BE}
			+\sqrt{p_{0_X}}\ket{2}_{A_1}\ket{1_{0_X}}_{BE}
			+\sqrt{p_{1_X}}\ket{3}_{A_1}\ket{1_{1_X}}_{BE},
		\end{split}
	\end{equation}
	and where we remark that $\ket*{n_a}_{BE}:=\ket*{n_a}_{B}\ket*{\tau}_{E}$.
	
	That is, if we define $\hat{\mathcal{D}}_{b}^{u}$ to be Bob-Eve's measurement operator associated to the outcome $b\in\set{0_Z,1_Z,0_X,1_X}$ in the $u$-th round, the probability of a phase error in that round for the reference virtual state can be written as
	\begin{equation}
		\Gamma_{1}^{u,{\rm ref}} =\bra{\Phi}\hat{\mathcal{D}}_{\text{ph},1}^u\ket{\Phi}_{ABCE}
		=p_{1}\left(p_Z^Bp_{0}^{\rm vir}\bra{1_{0}^{\rm vir}}\hat{\mathcal{D}}_{1_X}^{u}\ket{1_{0}^{\rm vir}}_{BE}+p_Z^Bp_{1}^{\rm vir}\bra{1_{1}^{\rm vir}}\hat{\mathcal{D}}_{0_X}^{u}\ket{1_{1}^{\rm vir}}_{BE}\right),
	\end{equation}
	where $\hat{\mathcal{D}}_{\text{ph},1}^u=\Pi_{1_C}\otimes(\ketbra{0_X}_{A_1}\otimes\hat{\mathcal{D}}_{1_X}^{u}+\ketbra{1_{X}}_{A_1}\otimes\hat{\mathcal{D}}_{0_X}^{u})$. On the other hand, we have that the probability of a single-photon bit error in the $X$ basis for the reference states can be written as
	\begin{equation}
		\Gamma_{X,1}^{u,{\rm ref}} =\bra{\Phi}\hat{\mathcal{D}}_{X,1}^u\ket{\Phi}_{ABCE} =p_{1}\left(p_X^Bp_{0_X}\bra{1_{0_X}}\hat{\mathcal{D}}_{1_X}^{u}\ket{1_{0_X}}_{BE}+p_X^Bp_{1_X}\bra{1_{1_X}}\hat{\mathcal{D}}_{0_X}^{u}\ket{1_{1_X}}_{BE}\right),
	\end{equation}
	where $\hat{\mathcal{D}}_{X,1}^u=\Pi_{1_C}\otimes(\ketbra{2}_{A_1}\otimes\hat{\mathcal{D}}_{1_X}^{u}+\ketbra{3}_{A_1}\otimes\hat{\mathcal{D}}_{0_X}^{u})$. Now we take advantage of the symmetries in the set of BB84 states (which imply that $\ket{1_0^{\rm vir}}=\ket{1_{0_X}}$ and $\ket{1_1^{\rm vir}}=\ket{1_{1_X}}$ according to \cref{eq:reference_virtual2}) to relate both errors as
	\begin{equation}
		\Gamma_{1}^{u,{\rm ref}}=\frac{p_Z^Bp_Z^A}{p_X^Bp_X^A}\Gamma_{X,1}^{u,{\rm ref}}.
	\end{equation}

	This means that one can estimate the number of phase errors from the sum of probabilities $\sum_{u=1}^N\Gamma_{X,1}^{u,{\rm ref}}$ in the reference framework by following a similar procedure to the previous section. We omit the details here for simplicity. In particular, we have that
	\begin{equation}\label{eq:Mph1}
		M_{\text{ph},1}\leq \bar{K}_{N,\epsilon}^{\rm U}\left(N G_{\delta}^{\rm U}\left(\frac{1}{N}\frac{p_Z^Bp_Z^A}{p_X^Bp_X^A}\sum_{u=1}^N\Gamma_{X,1}^{u,{\rm ref}}\right)\right).
	\end{equation}
	Now, the quantity $\sum_{u=1}^N\Gamma_{X,1}^{u,{\rm ref}}$ can be estimated by means of the decoy-state method to obtain an upper bound $M_{\rm ph,1}^{\rm U}\geq M_{\rm ph,1}$ from~\cref{eq:Mph1}. For this, note again that the reference states do not leak information about Alice's settings, and thus the probability $\Gamma_{X,1}^{u,{\rm ref}}$ can be bounded by a linear function of the observed bit-error statistics.
	
	We remark that the reference states are never sent in the real protocol, but the statistical relations between the mentioned quantities are still valid. In particular, one can upper bound $\sum_{u=1}^N\Gamma_{X,1}^{u,{\rm ref}}$ by solving the following LP:
	\begin{eqnarray*}
		\text{max} & \quad & \frac{1}{N}\sum_{u=1}^N\bar{\Gamma}^{u,{\rm ref}}_{X,1}\\
		\text{s.t.} & \quad & \frac{1}{Np_{\mu}p_X^Ap_X^B}\sum_{u=1}^N e^{u,{\rm ref}}_{X,\mu} -\Lambda_\mu \leq \sum_{n=0}^{n_{\text{cut}}}  p_{n|\mu} \left(\frac{1}{N}\sum_{u=1}^N\bar{\Gamma}^{u,{\rm ref}}_{X,n}\right)\leq \frac{1}{Np_{\mu}p_X^Ap_X^B}\sum_{u=1}^N e^{u,{\rm ref}}_{X,\mu}, \quad \forall \mu
	\end{eqnarray*}
	where $\bar{\Gamma}^{u,{\rm ref}}_{X,n}:=\Gamma_{X,n}^{u,{\rm ref}}/(p_np_X^Ap_X^B)$ is the conditional probability of observing a bit error in the $u$-th round given that Alice sent a $n$-photon state and both she and Bob select the $X$ basis, $e^{u,{\rm ref}}_{X,\mu}=\bra{\Phi}\hat{\mathcal{D}}_{X,\mu}^u\ket{\Phi}_{ABCE}$, with $\hat{\mathcal{D}}_{X,\mu}^u=\Pi_{\mu}\otimes(\ketbra{2}_{A_1}\otimes\hat{\mathcal{D}}_{1_X}^{u}+\ketbra{3}_{A_1}\otimes\hat{\mathcal{D}}_{0_X}^{u})$, is the probability that, in the reference scenario, Alice selects the intensity $\mu$, she and Bob select the $X$ basis, and a bit error occurs, conditioned on all the previous information announced by Alice and Bob up to the $u$-th round, and $\Lambda_\mu$ has been defined in~\cref{eq:Lambda}.
	
	Finally, the sum of probabilities $\sum_{u=1}^N e^{u,{\rm ref}}_{X,\mu}$ can be bounded, for each $\mu$, from the corresponding observed number of bit errors in the protocol, namely $E_{X,\mu}$, by following an analogous procedure to the previous subsection, which results in
	\begin{equation}
		G_{\delta}^{\rm L}\left(\frac{1}{N}K_{N,\epsilon}^{\rm L}(E_{X,\mu})\right)  \leq \frac{1}{N}\sum_{u=1}^N e^{u,{\rm ref}}_{X,\mu} \leq G_{\delta}^{\rm U}\left(\frac{1}{N}K_{N,\epsilon}^{\rm U}(E_{X,\mu})\right).
	\end{equation}

	\subsection{Secret-Key Rate}
	After obtaining the bounds $M_{\rm ph,1}^{\rm U}\geq M_{\rm ph,1}$ and $M_{1}^{Z, {\rm L}}\leq M_{1}^Z$, Alice and Bob perform error correction, error verification, and privacy amplification. The secret-key rate of the protocol is given by $R=l/N$ where, as shown in \cref{appendix:Security}, the length of the final key is given by
    \begin{equation}\label{eq:lenght}
        l = M_1^{Z,{\rm L}}\left(1-h(e_{\rm ph}^{\rm U})\right) - \lambda_{\rm EC} - \log_2\frac{1}{\epsilon_c} - 2\log_{2}\frac{1}{\epsilon_2}-1 - \log_2\frac{1}{4\epsilon_{\rm PA}},
    \end{equation}
	where $e_{\rm ph}^{\rm U}=M_{\rm ph,1}^{\rm U}/M_1^{Z,{\rm L}}$, and $\lambda_{\rm EC}$ is the number of bits revealed in the error correction process, which we set to $\lambda_{\rm EC}=M_Z f_e h(e_Z)$, \ie, it depends on the overall number of $Z$-basis detection events $M_Z$, the error-correction efficiency $f_e$, and the quantum bit error rate in the $Z$ basis $e_Z$. A detailed explanation of the meaning of all the remaining parameters $\epsilon_c$, $\epsilon_2$ and $\epsilon_{\rm PA}$ can be found in \cref{appendix:Security}.

	\section{Simulations}
	For the simulations, we use a typical channel model (see \cref{appendix:channelModel}) for a three-intensity decoy-state BB84 protocol. Furthermore, we fix the dark-count probability of Bob's detectors to $p_d=7.2\times 10^{-8}$ and their detection efficiency to $\eta_D=0.65$ (matching the parameters used in a recent experiment reported in~\cite{yin2016measurement}). Besides, we set the system misalignment to $\varphi_{\text{mis}}=6^{\circ}$, which roughly corresponds to an intrinsic error rate of 1\%, and we consider a typical fiber-loss coefficient $\alpha_{\si{\decibel}}=\SI{0.2}{\decibel\per\kilo\metre}$. Regarding the protocol parameters, for concreteness we set $\epsilon_{\rm s}=\epsilon_{\rm c}=10^{-10}$, and $f_e=1.2$, and, for simplicity, we impose $\epsilon_{\rm PA}=\epsilon_2=\epsilon_{\rm s}/3$, $\varepsilon=(\epsilon_{\rm s}/6)^2$, and $\epsilon=\varepsilon/14$. We note that the value of these latter probabilities can be chosen freely to maximize the secret-key rate, as long as they satisfy $\epsilon_{\rm s}=\epsilon_2+\epsilon_{\rm PA}+2\sqrt{\epsilon}$ (see~\cref{appendix:Security}).
	
	For each distance we optimize the two highest intensities $\mu_0$ and $\mu_1$, leaving the weakest intensity fixed to $\mu_2=10^{-4}$ due to the finite extinction ratio of real intensity modulators, which are the devices typically used to control the intensity of the transmitted pulses. Besides, we optimize the $Z$-basis selection probability, which we assume equal for Alice and Bob, \ie, $p_Z^A=p_Z^B$, and the probability that Alice selects the intensity $\mu_0$, namely $p_{\mu_0}$, being the remaining intensity probabilities fixed to $p_{\mu_1}=p_{\mu_2}=(1-p_{\mu_0})/2$ for simplicity.
	
	\begin{figure}[ht]
		\centering
		\includegraphics[width=0.65\textwidth]{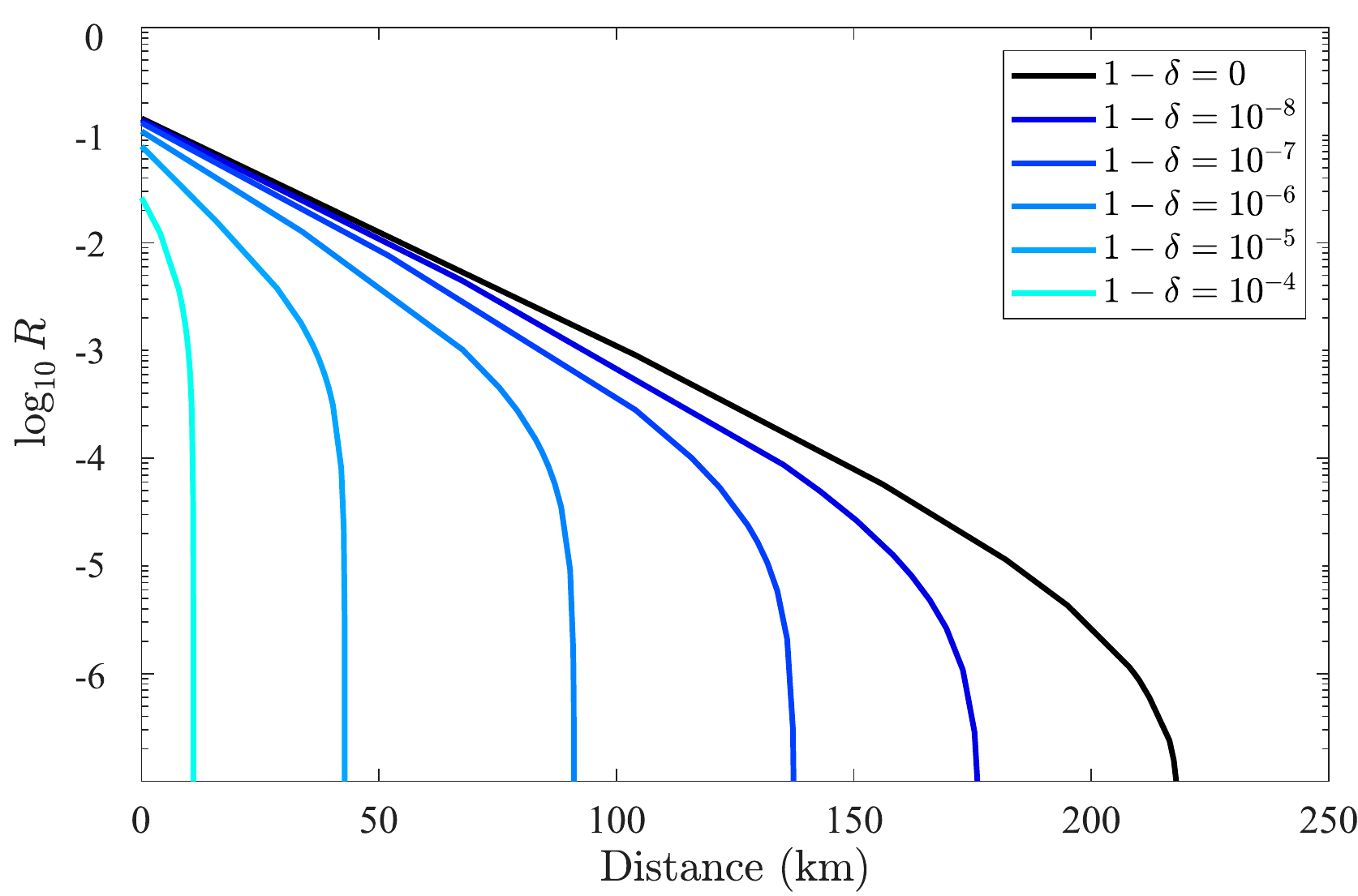}
		\caption{Secret-key rate $R$ in logarithmic scale for $N=10^{10}$ and for different values of the quantity $\delta$. The dark-count probability of Bob's detectors is set to $p_d=7.2\times 10^{-8}$, being their detection efficiency $\eta_D=0.65$ \cite{yin2016measurement}. Also, we fix the system's misalignment to $\varphi_{\text{mis}}=6^{\circ}$, and the loss coefficient of the channel to $\alpha_{\si{\decibel}}=\SI{0.2}{\decibel\per\kilo\metre}$. For further details, see the main text.} \label{fig:mainFig}
	\end{figure}

	The results are shown in \cref{fig:mainFig}, in which we evaluate the performance of the protocol for different values of $\delta$, being the number of transmitted signals equal to $N=10^{10}$ for all the curves. We find that the protocol allows to distill a secret key even for relatively low values of the parameter $\delta$, such as $\delta=1-10^{-4}$, but at the cost of reducing the maximum distance between the users. It is important to note that, in principle, the parameter $\delta$ can be made as close to 1 as desired by simply increasing the isolation of Alice's equipment. This is because the intensity of Eve's injected light could be limited in practice due to the laser-induced damage threshold, which provides an estimation of the maximum energy that can be injected into Alice's transmitter in a characteristic time interval without damaging it~\cite{lucamarini2015Practical}. Naturally, this quantity has to be measured experimentally, but once this experimental characterization is done, it provides a practical upper bound on the intensity of Eve's injected light. Given a certain value of the isolation, this results in an upper bound on the maximum intensity of Eve's back-reflected light, which can be used to lower bound $\delta$.
		
\begin{figure}[ht]
		\centering
		\includegraphics[width=0.65\textwidth]{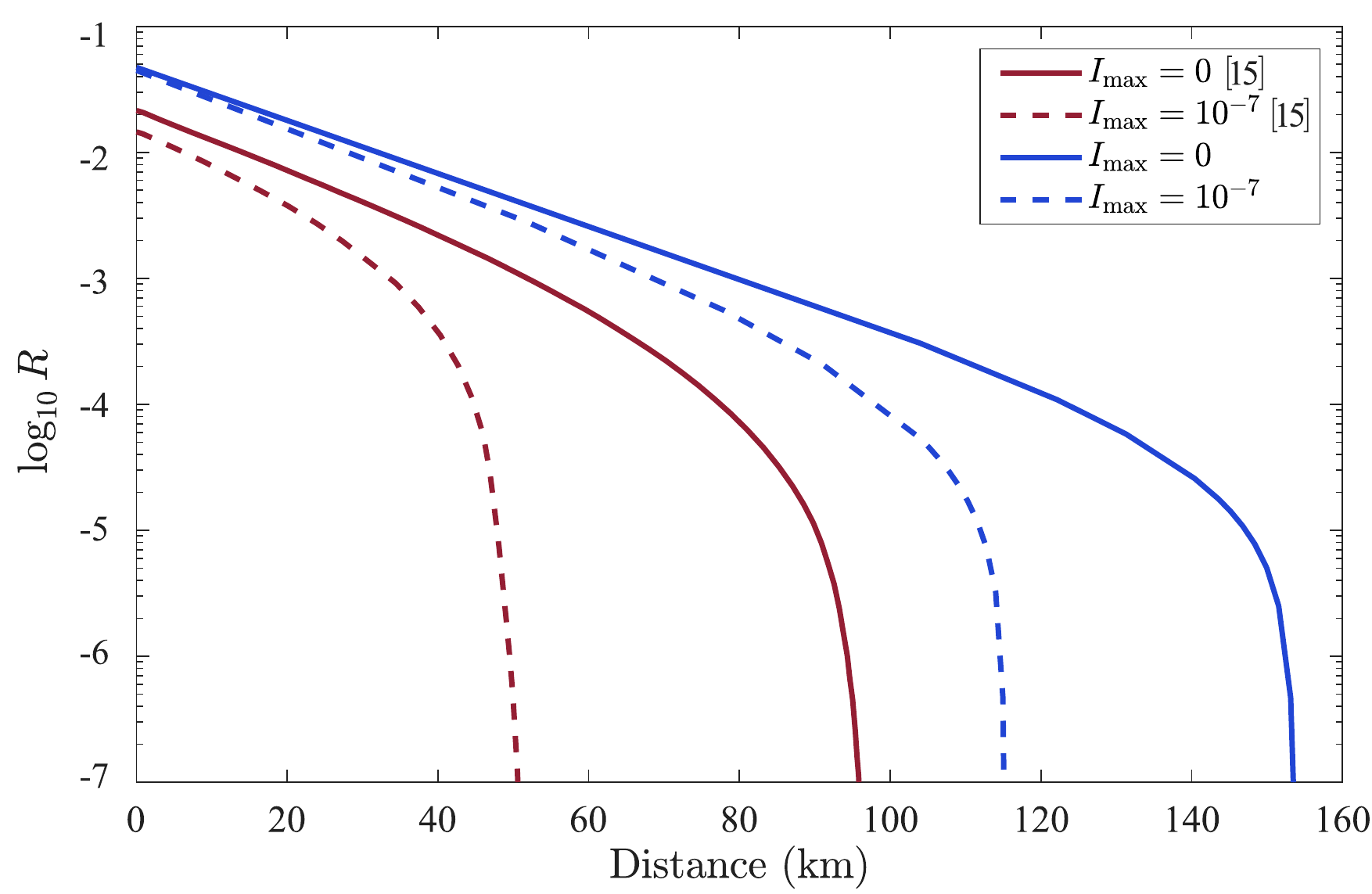}
		\caption{Comparison between the secret-key rates $R$, in logarithmic scale, obtained in the presence of a THA in which Eve injects strong coherent light~\cite{lucamarini2015Practical,tamaki2016decoy,wang2018finite} for the case of $N=10^{12}$ transmitted signals. The results associated to the security proof introduced in this paper are illustrated in blue, while those of Ref~\cite{wang2018finite}, which correspond to a finite key analysis of the asymptotic results in~\cite{lucamarini2015Practical,tamaki2016decoy}, are shown in magenta. In these simulations we use the same experimental parameters considered in~\cite{wang2018finite}, \ie,  $p_d=5\times 10^{-6}$, $\eta_D=0.25$, and an intrinsic error rate due to misalignment of 1\% (which roughly corresponds to a misalignment angle $\varphi_{\text{mis}}=6^{\circ}$).} \label{fig:coherentSimple}
	\end{figure}

	In \cref{fig:coherentSimple} we compare our results with the security proof introduced previously in~\cite{wang2018finite} based on the earlier works reported in~\cite{lucamarini2015Practical,tamaki2016decoy}. For this, we consider the case in which Eve's probe is a coherent state that does not modify the behavior of Alice's devices~\cite{lucamarini2015Practical,tamaki2016decoy,wang2018finite} and the back-reflected light leaks information about Alice's bit/basis and intensity settings. Specifically, we consider that such back-reflected light is a coherent state of the form $\ket{\beta_{a,\mu}e^{i\theta_{a,\mu}}}_E$, where $\beta_{a,\mu}^2\leq I_{\max}$ and the quantities $\beta_{a,\mu}$ and $\theta_{a,\mu}$ depend on Alice's setting choices $a$ and $\mu$~\cite{lucamarini2015Practical,tamaki2016decoy,wang2018finite}. This means that the state $\ket{\tilde{n}_{a,\mu}}_{BE}$ in \cref{eq:actual_virtual_state_amu} can be particularized here to $\ket{\tilde{n}_{a,\mu}}_{BE}=\ket{n_{a}}_{B}\otimes\ket{\beta_{a,\mu}e^{i\theta_{a,\mu}}}_E$, and the photon number statistics remain unaltered, \ie, $\tilde{p}_{n|\mu}=p_{n|\mu}$. Thus, according to~\cref{eq:delta1}, we have that, in this scenario, the parameter $\delta$ has the form
	\begin{equation}\label{eq:delta_wei}
	\begin{split}
	    \delta = \abs{\sum_{a,\mu}p_{a}p_{\mu}\sum_n p_{n|\mu}\braket{\tilde{n}_{a,\mu}}{n_{a}}_{BE}} =\abs{\sum_{a,\mu}p_{a}p_{\mu}\braket{\tau}{\beta_{a,\mu}e^{i\theta_{a,\mu}}}_{E}}.
	\end{split}
	\end{equation}
	If we set, for instance, $\ket{\tau}_E=\ket{\rm vac}_E$, which is a natural choice for the reference states if we assume that Eve's side-channel information is highly attenuated by Alice's isolator, we obtain
	\begin{equation}
		\delta=\abs{\sum_{a,\mu}p_{a}p_{\mu}e^{-\frac{\beta_{a,\mu}^2}{2}}}\geq e^{-\frac{I_{\max}}{2}}
	\end{equation}

	 For the numerical simulations we select the same experimental parameters considered in~\cite{wang2018finite}, which further assumes that the number of transmitted signals is $N=10^{12}$. In particular, the dark-count probability of Bob's detectors is now $p_d=5\times 10^{-6}$, their detection efficiency is $\eta_D=0.25$, and the intrinsic error rate due to misalignment is 1\% (which  for the channel model shown in \cref{appendix:channelModel} roughly corresponds to a misalignment angle $\varphi_{\text{mis}}=6^{\circ}$, as mentioned above). The improvement offered by the security proof introduced in this paper is rather remarkable, being now the maximum achievable distance more than twice of that obtained in~\cite{wang2018finite} for $I_{\max}=10^{-7}$. Indeed, it can be shown that, in terms of isolation, the security proof introduced in~\cite{wang2018finite} requires Alice to increase the isolation of her transmitter in roughly \SI{17}{\decibel} to achieve the same maximum distance that she could attain with the security proof presented in this work. Also, we note that our finite-key security analysis is much tighter than that in~\cite{wang2018finite}, as we can observe from the curves shown in~\cref{fig:coherentSimple} for the case of no information leakage, \ie, $I_{\max}=0$. 

	 In \cref{appendix:3stateLT} we provide the finite-key security analysis for a decoy-state-based three-state loss-tolerant (LT) protocol~\cite{tamaki2014loss}, and we compare its performance with the decoy-state BB84 scheme. We refer the readers to that appendix for the details. Finally, in~\cref{appendix:Intensity} we study the case in which the photon number statistics of Alice's signals might be partially modified by Eve, as has been demonstrated in~\cite{huang2019laser}.

	\section{Acknowledgements}
	This work was supported by the Galician Regional Government (consolidation of Research Units: AtlantTIC), the Spanish Ministry of Economy and Competitiveness (MINECO), the Fondo Europeo de Desarrollo Regional (FEDER) through Grant No. PID2020-118178RB-C21, and the Spanish Ministry of Science and Innovation through the “Planes Complementarios de I+D+I con las Comunidades Autónomas” in Quantum Communication.

	\section{Conclusions}
	In this work we have introduced a general finite-key security proof for decoy-state-based QKD in the presence of potential information leakages from Alice's transmitter, which could be produced, for instance, by a Trojan-horse attack (THA). For this, we have taken advantage of a Cauchy-Schwarz-based constraint to incorporate the information leakage from the bit/basis and intensity encoding setups in the security analysis. This constraint requires the users to bound a single parameter that encapsulates all the imperfections, and we have used novel concentration bounds to deal with the finite-key effects. In practice, such single parameter can be directly related to the amount of isolation of Alice's transmitter.
	
	For illustration purposes, we have evaluated the performance of the standard decoy-state BB84 protocol and the decoy-state loss-tolerant protocol in the presence of a THA. The results demonstrate the feasibility of both schemes over long distances given that the information leakage is small enough, which could be achieved by increasing the isolation of the devices. Our results significantly outperform previous approaches by doubling the maximum achievable distance in realistic scenarios.

	\appendix
	\section{Concentration bounds for dependent random variables}
	\label{appendix:ConcentrationBounds}
	Let $\xi_1,..., \xi_N$ be a sequence of Bernoulli random variables, and let $\Lambda_l = \sum_{u=1}^{l} \xi_u$. Let $\mathcal{F}_{l}$ be its natural filtration, \ie, the $\sigma$-algebra generated by $\{\xi_1,...,\xi_l\}$. According to Kato's inequality \cite{kato}, for any $N,a,b$ such that $b\geq \abs{a}$, we have that
	\begin{equation}
		\label{eq:katobound_upper}
		\begin{aligned}
			\Pr &\left[\sum_{u=1}^{N} \Pr(\xi_u = 1 \vert \mathcal{F}_{u-1}) - \Lambda_N  \geq \left[b+a \left(\frac{2 \Lambda_N}{N} - 1 \right) \right] \sqrt{N} \right] \leq \exp \left[\frac{-2(b^2-a^2)}{(1+\frac{4a}{3 \sqrt{N}})^2}  \right].
		\end{aligned}
	\end{equation}
	Besides, by replacing $\xi_l$ with $1-\xi_l$ and $a$ with $-a$ in \cref{eq:katobound_upper}, one obtains \cite{curras2021tight}
	\begin{equation}
		\label{eq:katobound_lower}
		\begin{aligned}
			\Pr &\left[\Lambda_N - \sum_{u=1}^{N} \Pr(\xi_u = 1 \vert \mathcal{F}_{u-1})  \geq \left[b+a \left(\frac{2 \Lambda_N}{N} - 1 \right) \right] \sqrt{N} \right] \leq \exp \left[\frac{-2(b^2-a^2)}{(1-\frac{4a}{3 \sqrt{N}})^2} \right].
		\end{aligned}
	\end{equation} 
	In~\cite{curras2021finite} it is shown how to use \cref{eq:katobound_upper} to derive an upper bound on the sum of conditional probabilities, namely $\sum_{u=1}^{N} \Pr(\xi_u = 1 \vert \mathcal{F}_{u-1})\leq K_{N,\epsilon}^{\rm U}\left(\Lambda_N\right)$, and \cref{eq:katobound_lower} to derive the corresponding lower bound $\sum_{u=1}^{N} \Pr(\xi_u = 1 \vert \mathcal{F}_{u-1})\geq K_{N,\epsilon}^{\rm L}\left(\Lambda_N\right)$ together with an upper bound on the actual number $\Lambda_N$, namely $\Lambda_N\leq \bar{K}_{N,\epsilon}^{\rm U}\left(\sum_{u=1}^{N} \Pr(\xi_u = 1 \vert \mathcal{F}_{u-1})\right)$. Here we show how to derive the remaining lower bound $\Lambda_N\geq \bar{K}_{N,\epsilon}^{\rm L}\left(\sum_{u=1}^{N} \Pr(\xi_u = 1 \vert \mathcal{F}_{u-1})\right)$ from~\cref{eq:katobound_upper}. 
	
	For this, let us assume that we know an upper bound $S$ on the sum of probabilities $\sum_{u=1}^{N} \Pr \left(\xi_u = 1 \vert \mathcal{F}_{u-1}\right)$. Before running the protocol one should use the previous knowledge to come up with a prediction $\tilde{S}$ of the value of $S$. Then, one calculates the values of $a$ and $b$ that yield to the tightest bound from \cref{eq:katobound_upper} if the prediction comes true. These values can be found by solving the following optimization problem
	\begin{equation}
		\begin{aligned}
			&\max_{a,b} & \frac{N}{\sqrt{N}+2a} \left(\frac{1}{\sqrt{N}} \tilde{S} +a -b  \right) \\
			&\textrm{s.t. }  &\exp \left[\frac{-2(b^2-a^2)}{(1+\frac{4a}{3 \sqrt{N}})^2} \right] = \epsilon, \\
			&& b \geq \abs{a},
		\end{aligned}
	\end{equation}
	whose analytical solution is\footnote{In some regimes (corresponding to particular combinations of the parameters $N$, $\tilde{S}$ and $\epsilon$) that are not relevant for the scenario considered in this paper, the values of $a$ and $b$ given in~\cref{eq:optimal_ab} might not yield to the optimal solution. We remark, however, that any values of $a$ and $b$ yield a valid bound as long as they satisfy the condition $b \geq \abs{a}$.}
	\begin{equation}\label{eq:optimal_ab}
		\begin{gathered}
			a=\frac{3 \sqrt{N} \left(-9 \left(3 N^2-8 N \tilde{S}+8 \tilde{S}^2\right) \ln\epsilon+9 \sqrt{N (N-2 \tilde{S})^2 \ln\epsilon (N \ln\epsilon-18 \tilde{S} (N-\tilde{S}))}-4 N \ln^2\epsilon\right)}{4 \left(36 \left(N^2-2 N \tilde{S}+2 \tilde{S}^2\right) \ln\epsilon+81 N \tilde{S} (N-\tilde{S})+4 N \ln^2\epsilon \right)}, \\
			b=\frac{1}{3} \sqrt{9 a^2-\frac{\left(4 a+3 \sqrt{N}\right)^2 \ln\epsilon }{2 N}}.
		\end{gathered}
	\end{equation}
	Then, we have that,
	\begin{equation}
		\label{eq:katobound_inverse_final}
		\bar{K}_{N,\epsilon}^{\rm L}(S):=\frac{N}{\sqrt{N}+2a} \left(\frac{1}{\sqrt{N}} S +a -b\right)\leq\Lambda_N,
	\end{equation}
	except with probability $\epsilon$. 
	
	Below we include for completeness the remaining bounds that we use in this work~\cite{curras2021tight,curras2021finite}, being all of them held except with probability $\epsilon$. In particular, an upper bound on the actual value $\Lambda_N$ is given by
	\begin{equation}
		\label{eq:katobound_inverse_final_UB}
		\bar{K}_{N,\epsilon}^{\rm U}(S):=\frac{N}{\sqrt{N}-2a} \left(\frac{1}{\sqrt{N}} S -a +b\right)\geq \Lambda_N,
	\end{equation}
	where
	\begin{equation}
		\begin{gathered}
			a = \frac{3 \sqrt{N} \left(9 \left(3 N^2-8 N \tilde{S}+8 \tilde{S}^2\right)\ln\epsilon+9 (N-2 \tilde{S}) \sqrt{N \ln \epsilon  (N \ln \epsilon +18 \tilde{S} (\tilde{S}-N))}+4 N \ln ^2\epsilon \right)}{4
				\left(36 \ln \epsilon  \left(N^2-2 N \tilde{S}+2 \tilde{S}^2\right)+4 N \ln ^2\epsilon +81 N \tilde{S} (N-\tilde{S})\right)}, \\
			b = \frac{\sqrt{18 a^2 N-\left(16 a^2-24 a \sqrt{N}+9 N\right) \ln\epsilon}}{3 \sqrt{2N}}.
		\end{gathered}
	\end{equation}

	A lower bound on the sum $S$ is given by
	\begin{equation}\label{eq:katobound_final_LB}
		K_{N,\epsilon}^{\rm L}(\Lambda_N):=\Lambda_N-\left[b+a \left(\frac{2 \Lambda_N}{N} - 1 \right) \right] \sqrt{N} \leq S,
	\end{equation}
	where the optimal values for $a$ and $b$ can be obtained if we have a prediction of $\Lambda_N$, which we denote $\tilde{\Lambda}_N$. To obtain this prediction, one could use data from previous executions of the protocol or a theoretical model for the quantum channel. Based on this prediction, the optimal values for $a$ and $b$ are given by
	\begin{equation}
		\begin{gathered}
			a = \frac{3 \left(-72 \sqrt{N} \tilde{\Lambda}_N  (N-\tilde{\Lambda}_N)\ln\epsilon + 16 N^{3/2} \ln^2\epsilon+9 \sqrt{2} (N-2 \tilde{\Lambda}_N) \sqrt{-N^2 \ln \epsilon  (9 \tilde{\Lambda}_N (N-\tilde{\Lambda}_N)-2 N \ln \epsilon)}\right)}{4 (9 N-8\ln
				\epsilon) (9 \tilde{\Lambda}_N (N-\tilde{\Lambda}_N)-2 N \ln\epsilon)}, \\
			b = \frac{\sqrt{18 a^2 N-\left(16 a^2+24 a \sqrt{N}+9 N\right) \ln\epsilon}}{3 \sqrt{2N}}.
		\end{gathered}
	\end{equation}

	Finally, an upper bound on the sum $S$ is given by
	\begin{equation}\label{eq:katobound_final_UB}
		K_{N,\epsilon}^{\rm U}(\Lambda_N):=\Lambda_N+\left[b+a \left(\frac{2 \Lambda_N}{N} - 1 \right) \right] \sqrt{N} \geq S,
	\end{equation}
	where
	\begin{equation}
		\begin{gathered}
			a = \frac{3 \left(72 \sqrt{N} \tilde{\Lambda}_N  (N-\tilde{\Lambda}_N)\ln \epsilon-16 N^{3/2} \ln^2\epsilon+9 \sqrt{2} (N-2 \tilde{\Lambda}_N) \sqrt{-N^2 \ln \epsilon  (9 \tilde{\Lambda}_N (N-\tilde{\Lambda}_N)-2 N \ln \epsilon)}\right)}{4 (9 N-8\ln
				\epsilon) (9 \tilde{\Lambda}_N (N-\tilde{\Lambda}_N)-2 N \ln\epsilon)}, \\
			b = \frac{\sqrt{18 a^2 N-\left(16 a^2+24 a \sqrt{N}+9 N\right) \ln\epsilon}}{3 \sqrt{2N}}.
		\end{gathered}
	\end{equation}
	%

	\section{Secrecy analysis}\label{appendix:Security}
	The derivation of the secret-key length given by~\cref{eq:lenght} is mainly based on~\cite{lim2014concise,tomamichel2012tight,curras2021tight}. Precisely, let \textbf{Z} ($\textbf{Z}'$) be Alice's (Bob's) sifted key of length $l_{\rm sif}$, and let $\textbf{E}'$ denote Eve's side information after the error correction step in which Bob's key $\textbf{Z}'$ becomes a copy of \textbf{Z} with very high probability. The Quantum Leftover Hash Lemma~\cite{tomamichel2011leftover} states that it is possible to extract a $\epsilon_{\rm s}$-secret key of length $l$ by applying privacy amplification with a random universal$_2$ hash function to $\textbf{Z}$. In particular, it says that, for any $\epsilon>0$,
	\begin{equation}\label{eq:QLHL}
	    \epsilon_{\rm s} \leq \epsilon + \frac{1}{2}\sqrt{2^{l-H_{\min}^{\epsilon}(\textbf{Z}|\textbf{E}')_{\rho}}},
	\end{equation}
	where $H_{\min}^{\epsilon}(\textbf{Z}|\textbf{E}')_{\rho}$ is the $\epsilon$-smooth min entropy of \textbf{Z} conditioned on $\textbf{E}'$, and $\rho$ is the quantum state that Alice measures to obtain \textbf{Z}. Roughly speaking, this means that if Alice and Bob can lower bound Eve's uncertainty about \textbf{Z} for a particular $\epsilon$, then they could choose $l$ to be maximal while satisfying~\cref{eq:QLHL}. Thus, in order to lower bound Eve's information, we first apply the chain rule for smooth min-entropies, which states~\cite{tomamichel2012tight}
	\begin{equation}\label{eq:error_correction_entropies}
	    H^{\varepsilon}_{\min}(\textbf{Z}|\textbf{E}')_{\rho} \geq H^{\varepsilon}_{\min}(\textbf{Z}|\textbf{E})_{\rho} - \lambda_{\rm EC} - \log_2\frac{1}{\epsilon_c},
	\end{equation}
	where $\lambda_{\rm EC}$ ($\log_2\frac{1}{\epsilon_c}$) is the number of bits revealed in the error correction (verification) step of the protocol, and \textbf{E} denotes Eve's information before the error correction step.

    Now we decompose \textbf{Z} into $\textbf{Z}_1\textbf{Z}_{\text{rest}}$, where $\textbf{Z}_1$ contains those bits of \textbf{Z} corresponding to single-photon events, and $\textbf{Z}_{\text{rest}}$ contains the remaining bits. By using the generalized chain rule from~\cite{vitanov2013chain} we have that
    \begin{equation}\label{eq:decoy_entropies}
    \begin{split}
        H^{\varepsilon}_{\min}(\textbf{Z}|\textbf{E})_{\rho} & \geq  H^{\epsilon_1}_{\min}(\textbf{Z}_1|\textbf{Z}_{\text{rest}}\textbf{E})_{\rho} + H^{\epsilon_3}_{\min}(\textbf{Z}_{\text{rest}}|\textbf{E})_{\rho} - 2\log_{2}\frac{1}{\epsilon_2}-1\\
        & \geq  H^{\epsilon_1}_{\min}(\textbf{Z}_1|\textbf{Z}_{\text{rest}}\textbf{E})_{\rho} - 2\log_{2}\frac{1}{\epsilon_2}-1,
    \end{split}
    \end{equation}
    where $\varepsilon=2\epsilon_1+\epsilon_2+\epsilon_3$, $\epsilon_2>0$, and $\epsilon_1,\epsilon_3\geq0$. In the second inequality we use $H^{\epsilon_3}_{\min}(\textbf{Z}_{\text{rest}}|\textbf{E})_{\rho}\geq0$. 
    
    Thanks to the previous step, we can focus now on bounding Eve's information about the single-photon events. In order to bound this quantity, we make use of the uncertainty relation for smooth entropies~\cite{tomamichel2011uncertainty}. For this, let $\textbf{X}_1$ be the outcome of Alice's measurement in the virtual scenario in which she measures all her ancillas associated to the single-photon signals in the complementary basis $X$. Thus, we have
    \begin{equation}\label{eq:uncertainty_rels}
    \begin{split}
        H^{\epsilon_1}_{\min}(\textbf{Z}_1|\textbf{Z}_{\text{rest}}\textbf{E})_{\rho} & \geq M_{1}^Z - H^{\epsilon_1}_{\max}(\textbf{X}_1|\textbf{B})_{\rho}\\        
        & \geq M_{1}^Z - H^{\epsilon_1}_{\max}(\textbf{X}_1|\textbf{X}_1')_{\rho}
    \end{split}
    \end{equation}
    where $\textbf{X}_1'$ is the bit string Bob would obtain if he measured system \textbf{B} in the complementary basis, and $H^{\epsilon}_{\max}(\cdot)_{\rho}$ is the $\epsilon$-smooth max entropy function. The second inequality comes from $H^{\epsilon_1}_{\max}(\textbf{X}_1|\textbf{X}_1')_{\rho} \geq H^{\epsilon_1}_{\max}(\textbf{X}_1|\textbf{B})_{\rho}$, since \textbf{B} cannot contain less information than $\textbf{X}_1'$.

    In the protocol, based on the result $k$ of all the measurements made by Alice and Bob to test the channel, they estimate a particular upper bound $e_{\rm ph}^{\rm U}(k)=M_{\rm ph,1}(k)/M_{1}^Z(k)$. Let\footnote{Note that the probability $\varepsilon(k)$, which is conditioned on the test outcomes, is unknown to the users. In the main test we consider the \textit{a priori} probability $\Pr(e_{\rm ph}>e_{\rm ph}^{\rm U})$.} $\varepsilon(k)=\Pr(e_{\rm ph}>e_{\rm ph}^{\rm U}(k) | k)$, being $e_{\rm ph}$ the fraction of bits that differ between $\textbf{X}_1$ and $\textbf{X}_1'$. From~\cite{tomamichel2012tight} we have that
    \begin{equation}\label{eq:tom2012}
        H_{\max}^{\sqrt{\epsilon(k)}}(\textbf{X}_1|\textbf{X}_1')_{\rho(k)} \leq M_{1}^Zh(e_{\rm ph}^{\rm U}(k)),
    \end{equation}
    and combining~\cref{eq:error_correction_entropies,eq:decoy_entropies,eq:uncertainty_rels,eq:tom2012} we have that, given $\Pr(e_{\rm ph}>e_{\rm ph}^{\rm U}(k) | k) = \varepsilon(k)$, the following is satisfied
    \begin{equation}
       H^{\varepsilon}_{\min}(\textbf{Z}|\textbf{E}')_{\rho(k)} \geq M_{1}^Z\left(1-h(e_{\rm ph}^{\rm U}(k))\right) - \lambda_{\rm EC} - \log_2\frac{1}{\epsilon_c} - 2\log_{2}\frac{1}{\epsilon_2}-1
    \end{equation}
    where $\varepsilon=2\sqrt{\varepsilon(k)}+\epsilon_2$. Finally, if we define $\epsilon_{\rm PA}:=\frac{1}{2}\sqrt{2^{l-H_{\min}^{\epsilon}(\textbf{Z}|\textbf{E}')_{\rho}}}$ and we substitute $\varepsilon$ with $\sqrt{\varepsilon(k)}+\epsilon_2$ in that expression and in~\cref{eq:QLHL}, we find that it is possible to extract a $\epsilon_{\rm s}$-secret key of length
    \begin{equation}
        l = M_{1}^Z\left(1-h(e_{\rm ph}^{\rm U}(k))\right) - \lambda_{\rm EC} - \log_2\frac{1}{\epsilon_c} - 2\log_{2}\frac{1}{\epsilon_2}-1 - \log_2\frac{1}{4\epsilon_{\rm PA}},
    \end{equation}
    satisfying $2\sqrt{\varepsilon(k)}+\epsilon_2+\epsilon_{\rm PA} \geq \epsilon_{\rm s}(k)$. Since we are interested in the overall secrecy parameter $\epsilon_{\rm s}$, we can bound it from
    \begin{equation}
        \epsilon_{\rm s} = \sum_k p(k)\epsilon_{\rm s}(k) 
        \leq \epsilon_2+\epsilon_{\rm PA} +2\sum_kp(k)\sqrt{\varepsilon(k)} 
        \leq \epsilon_2+\epsilon_{\rm PA} +2\sqrt{\sum_kp(k)\varepsilon(k)}  
        \leq \epsilon_2+\epsilon_{\rm PA}+2\sqrt{\varepsilon}
    \end{equation}
    where in the second inequality we have applied Jensen's inequality, and in the last inequality we have applied $\sum_kp(k)\varepsilon(k)=\Pr(e_{\rm ph}>e_{\rm ph}^{\rm U})\leq \varepsilon$, which is proven in the main text.

	\section{Channel model}\label{appendix:channelModel}
\begin{figure}[ht]
		\centering
		\includegraphics[width=0.65\textwidth]{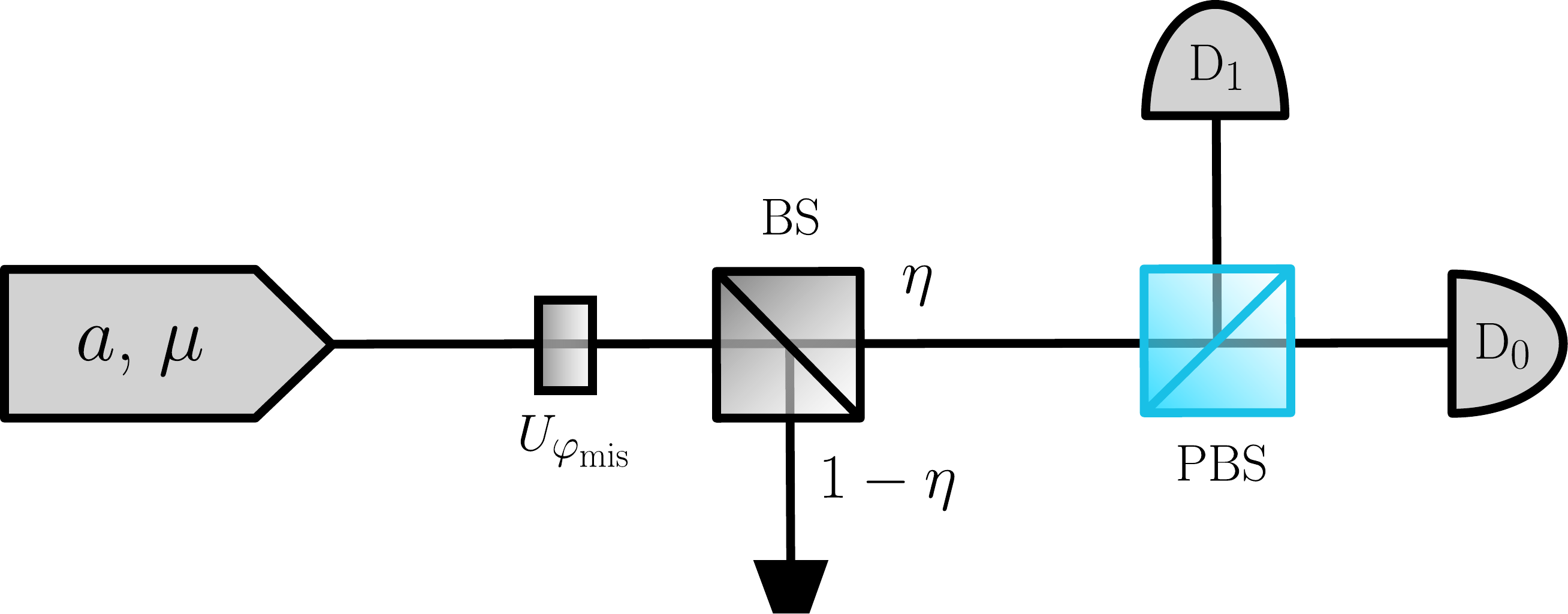}
		\caption{Schematic representation of the channel model. Alice selects the settings $a$ and $\mu$ and prepares a weak coherent pulse with intensity $\mu$ and polarization angle $\varphi_a$. We recall that $\varphi_{\rm mis}$ ($\eta$) stands for the misalignment introduced by de channel (overall system efficiency). In the figure, we consider for simplicity a fixed $Z$-basis measurement at Bob's side, which consists in a polarizing beamsplitter (PBS) and two threshold single-photon detectors (D$_0$ and D$_1$).} \label{fig:channel}
	\end{figure}

	We consider a typical channel model for decoy-state QKD based on polarization encoding, which is sketched in \cref{fig:channel}. In particular, for each value of $a\in\set{0_Z,1_Z,0_X,1_X}$, Alice prepares a coherent state with polarization angle $\varphi_a\in\set{0,\frac{\pi}{2},\frac{\pi}{4},\frac{3\pi}{4}}$. Besides, she sets the amplitude of each transmitted pulse accordingly to the intensity setting $\mu$. Note that in this model we are not required to consider a random phase for Alice's coherent states since we are describing an honest implementation of the quantum channel in which the phases of the coherent states do not play any role. That is, we would obtain exactly the same result if we considered phase-randomized coherent states in the calculations below.

	In the quantum channel, we model the polarization misalignment with a unitary operation $U_{\varphi_{\rm mis}}$ that makes the creation operators of its input modes evolve as $\hat{a}_{\rm H}^{\dagger}\to \cos\varphi_{\rm mis}\hat{b}_{\rm H}^{\dagger}+\sin\varphi_{\rm mis}\hat{b}_{\rm V}^{\dagger}$ and $\hat{a}_{\rm V}^{\dagger}\to \cos\varphi_{\rm mis}\hat{b}_{\rm V}^{\dagger}-\sin\varphi_{\rm mis}\hat{b}_{\rm H}^{\dagger}$, where $\hat{a}_{\rm H}^{\dagger}$ and $\hat{a}_{\rm V}^{\dagger}$ ($\hat{b}_{\rm H}^{\dagger}$ and $\hat{b}_{\rm V}^{\dagger}$) are, respectively, the creation operators associated to the horizontal and vertical polarization modes before (after) the misalignment. On the other hand, the overall system efficiency is modeled with a beamsplitter (BS) of transmittance $\eta=\eta_D\eta_c$, where $\eta_D$ is the efficiency of Bob's detectors, $\eta_c=10^{-\alpha_{\si{\decibel}}L/10}$ is the transmittance of the quantum channel, $\alpha_{\si{\decibel}}$ is the fiber-loss coefficient, and $L$ is the total distance between Alice and Bob. 
	
	Finally, at Bob's side the horizontal and vertical modes are spatially separated with a polarizing beamsplitter (PBS) whose output ports are connected to two threshold single-photon detectors of perfect efficiency (since $\eta_D$ has already been considered in $\eta$) and dark-count probability $p_d$.
	
	For this simple model, the \textit{click} probability at Bob's side given that Alice prepares a signal with intensity $\mu$ is given by
	\begin{equation}
		Q_{\mu}=1-(1-p_d)^2e^{-\mu\eta},
	\end{equation}
	which means that the number of $Z$-basis detections in which Alice selects the intensity $\mu$ satisfies $M_{\mu}^Z\approx Np_{Z}^Ap_{Z}^Bp_{\mu}Q_{\mu}$. On the other hand, it can be shown that the probability $P_{b|a,\mu,Z_B}$ that Bob observes a particular outcome $b\in\set{0,1}$ when he measures the incoming signal in the $Z$ basis given that Alice selected the settings $a$ and $\mu$ is
	\begin{equation}\label{eq:channel_model_P}
		\begin{split}
			P_{0_Z|a,\mu,Z_B} &= (1-p_d)\left(e^{-\eta\mu\sin^2\varphi_a'} - (1-p_d) e^{-\eta\mu}\right) + \frac{1}{2}h_{\varphi_a'},\\
			P_{1_Z|a,\mu,Z_B} &= (1-p_d)\left(e^{-\eta\mu\cos^2\varphi_a'} - (1-p_d) e^{-\eta\mu}\right) + \frac{1}{2}h_{\varphi_a'},
		\end{split}
	\end{equation}
	where
	\begin{equation}\label{eq:channel_model_h}
		\begin{split}
			h_{\varphi_a'} &= \left(1-e^{-\eta\mu\sin\varphi_a'}\right)\left(1-e^{-\eta\mu\cos\varphi_a'}\right) + p_d\left(e^{-\eta\mu\sin\varphi_a'}+e^{-\eta\mu\cos\varphi_a'}-2e^{-\eta\mu}\right)+p_d^2e^{-\eta\mu},
		\end{split}
	\end{equation}
	and $\varphi_a'=\varphi_a+\varphi_{\rm mis}$. In~\cref{eq:channel_model_P,eq:channel_model_h} we have considered that each double-\textit{click} event at Bob's side is randomly re-assigned to a single-\textit{click} event in one of the two detectors D$_0$ or D$_1$ (see \cref{fig:channel}). Also, note that due to the symmetry of the model, we have that $P_{b|a,\mu,X_B}$ can be obtained from $P_{b|a,\mu,Z_B}$ by simply shifting the angle $\varphi_a$ by $\pi/2$ radians. With this, we can write $M_{a,b_Z,\mu}\approx Np_{\mu}p_ap_{Z}^BP_{b|a,\mu,Z_B}$ and $M_{a,b_X,\mu}\approx Np_{\mu}p_ap_{X}^BP_{b|a,\mu,X_B}$.
	
	Finally, from the previous expressions, we find that the bit error probability in the $Z$ basis given that Alice selects the intensity $\mu$ can be written as
	\begin{equation}
		e_{Z,\mu}=\frac{P_{1_Z|0_Z,\mu,Z_B}+P_{0_Z|1_Z,\mu,Z_B}}{2},
	\end{equation}
	which, in the model above, is exactly equal to the corresponding error probability in the $X$ basis, namely $e_{X,\mu}$. This means that $E_{X,\mu}\approx Np_{X}^Ap_{X}^Bp_{\mu}e_{X,\mu}$.

	\section{Decoy-state three-state QKD protocol}\label{appendix:3stateLT}
	\subsection{Estimation of the number of phase errors}
	Here we consider that Alice and Bob implement the loss-tolerant (LT) three-state QKD protocol introduced in~\cite{tamaki2014loss} with phase-randomized weak coherent pulses and decoy states. The estimation of $M_1^Z$ is completely analogous to that of the decoy-state BB84 protocol, so below we focus on the estimation of the number of phase errors. The definition of the virtual state is also analogous to that of the BB84 case, but now with $a\in\set{0_Z,1_Z,0_X}$. For convenience, we consider an equivalent virtual scenario in which Alice prepares in each round the following virtual state
	\begin{equation}\label{eq:actual_virtual_separated}
		\ket{\Psi}_{ABCE}=\sum_{a,\mu}\sqrt{p_{a}p_{\mu}\tilde{p}_{n|\mu}}\ket{R_a,R_{\mu}}_{A}\sum_{n\neq1}\ket{n}_{C}\ket{\tilde{n}_{a,\mu}}_{BE} +\sum_{\mu}\sqrt{p_{\mu}\tilde{p}_{1|\mu}}\ket{R_{\mu}}_{A_2}\ket{1}_{C}\ket{\psi_{\mu,1}}_{A_1BE},
	\end{equation}
	with
	\begin{equation}
		\begin{split}\label{eq:actual_virtual_singleLT}
			\ket{\psi_{\mu,1}}_{A_1BE} = & \sqrt{p_Z^Bp_{0_Z}}\ket{0}_{A_1}\ket{\tilde{1}_{0_Z}}_{BE} +\sqrt{p_Z^Bp_{1_Z}}\ket{1}_{A_1}\ket{\tilde{1}_{1_Z}}_{BE}
			+\sqrt{p_Z^Bp_{0_X}}\ket{2}_{A_1}\ket{\tilde{1}_{0_X}}_{BE}\\
			& +\sqrt{p_X^Bp_{0_Z}}\ket{3}_{A_1}\ket{\tilde{1}_{0_Z}}_{BE} +\sqrt{p_X^Bp_{1_Z}}\ket{4}_{A_1}\ket{\tilde{1}_{1_Z}}_{BE}
			+\sqrt{p_X^Bp_{0_X}}\ket{5}_{A_1}\ket{\tilde{1}_{0_X}}_{BE}.
		\end{split}
	\end{equation}
	That is, when Alice observes $\ketbra{1}_C$, she prepares the virtual state given by \cref{eq:actual_virtual_singleLT}.
	Specifically, in the virtual scenario Alice and Bob jointly measure system $A_1$ in the $\set{\ket{0_X},\ket{1_X},\ket{2},\ket{3},\ket{4},\ket{5}}$ basis, and subsequently Bob measures system $B$ accordingly to the outcome of $A_1$. That is, similar to the BB84 case, the probability of a phase error in the $u$-th round is given by $\Gamma_{1}^{u}=\bra*{\Psi}\hat{\mathcal{D}}_{\text{ph},1}^u\ket*{\Psi}_{ABCE}$, where $\hat{\mathcal{D}}_{\text{ph},1}^u=\Pi_{1_C}\otimes(\ketbra{0_X}_{A_1}\otimes\hat{\mathcal{D}}_{1_X}^{u}+\ketbra{1_{X}}_{A_1}\otimes\hat{\mathcal{D}}_{0_X}^{u}).$
	
	Now, following an analogous procedure to the first two steps described in \cref{subsec:single-photon-yield} we find an upper bound on the number of single-photon phase errors, which is
	\begin{equation}
		M_{\text{ph},1}\leq K_{N,\epsilon}^{\rm U}\left(N G_{\delta}^{\rm U}\left(\frac{1}{N}\sum_{u=1}^N\Gamma_{1}^{u,{\rm ref}}\right)\right),
	\end{equation}
	where $\Gamma_{1}^{u,{\rm ref}}=\bra{\Phi}\hat{\mathcal{D}}_{\rm ph}^u\ket{\Phi}_{ABCE}$ is the single-photon phase-error probability for the reference state $\ket{\Phi}_{ABCE}$ conditioned on all the previous information available up to the $u$-th round. Besides, we can write the reference state in terms of the complementary basis, \ie,
	\begin{equation}\label{eq:reference_virtual_separated}
		\ket{\Phi}_{ABCE}=\sum_{a,\mu}\sqrt{p_{a}p_{\mu}p_{n|\mu}}\ket{R_a,R_{\mu}}_{A}\sum_{n\neq1}\ket{n}_{C}\ket{n_{a}}_{BE} +\sum_{\mu}\sqrt{p_{\mu}p_{1|\mu}}\ket{R_{\mu}}_{A_2}\ket{1}_{C}\ket{\phi_{1}}_{A_1BE},
	\end{equation}
	where
	\begin{equation}\label{eq:phi_ref_mu_1}
		\begin{split}
			\ket{\phi_{1}}_{A_1BE} =& \sqrt{p_Z^Bp_{0}^{\rm vir}}\ket{0_X}_{A}\ket{1_{0}^{\rm vir}}_{BE} +\sqrt{p_Z^Bp_{1}^{\rm vir}}\ket{1_X}_{A}\ket{1_{1}^{\rm vir}}_{BE}
			+\sqrt{p_Z^Bp_{0_X}}\ket{2}_{A}\ket{1_{0_X}}_{BE}\\
			& +\sqrt{p_X^Bp_{0_Z}}\ket{3}_{A}\ket{1_{0_Z}}_{BE}
			 +\sqrt{p_X^Bp_{1_Z}}\ket{4}_{A}\ket{1_{1_Z}}_{BE}
			+\sqrt{p_X^Bp_{0_X}}\ket{5}_{A}\ket{1_{0_X}}_{BE}.
		\end{split}  
	\end{equation}

	With the previous definition, and for the particular set of states $\set{\ket{1_{0_Z}},\ket{1_{1_Z}},\ket{1_{0_X}}}=\set{\ket{0_Z},\ket{1_Z},\ket{0_X}}$, it is easy to show that $\Gamma^{u,{\rm ref}}_1$ can be written as
	\begin{equation}\label{eq:LT_error_rel}
		\Gamma^{u,{\rm ref}}_1=\frac{p_Z^Bp_{0}^{\rm vir}}{p_X^Bp_{0_X}}Y_{5,1_X,1}^{u,\rm ref} + p_Z^Bp_{1}^{\rm vir}\left(\frac{Y_{3,0_X,1}^{u,\rm ref}}{p_X^Bp_{0_Z}} +\frac{Y_{4,0_X,1}^{u,\rm ref}}{p_X^Bp_{1_Z}} -\frac{Y_{5,0_X,1}^{u,\rm ref}}{p_X^Bp_{0_X}}\right),
	\end{equation}
	where $Y_{a,b,n}^{u,\rm ref}:=\bra{\Phi}\Pi_{n_C}\otimes\Pi_{a}\otimes\hat{\mathcal{D}}_{b}^u\ket{\Phi}_{ABCE}$ is the probability that Alice selects the setting $a$, sends a $n$-photon pulse, and Bob observes the outcome $b\in\set{0_Z,1_Z,0_X,1_X}$ conditioned on all the information available up to the $u$-th round, and where $\Pi_{a}=\ketbra{R_a}_{A_{1}}$. We note that $\Gamma^{u,{\rm ref}}_1$ can always be written as a function of the yields $Y_{a,b,n}^{u,{\rm ref}}$ independently on the set of reference states that we choose (\ie, even if we consider the case of flawed states) given that they lie in a qubit space~\cite{tamaki2014loss,curras2021finite}.
	
	 Next, we take the sum over all rounds in both sides of the equality given in~\cref{eq:LT_error_rel} and estimate each quantity $\sum_{u=1}^N Y_{a,b,n}^{u,{\rm ref}}$ through the decoy-state method. That is, we solve, for instance, the following LP:
	\begin{eqnarray*}
		\text{min/max} & \quad & \frac{1}{N}\sum_{u=1}^N \tilde{Y}_{a,b,n}^{u,{\rm ref}}:=\frac{1}{Np_{n}p_ap_X^B}\sum_{u=1}^N Y_{a,b,n}^{u,{\rm ref}}\\
		\text{s.t.} & \quad & \frac{1}{p_ap_X^Bp_{\mu}N}\sum_{u=1}^N Q^{u,{\rm ref}}_{a,b,\mu} -\Lambda_{\mu} \leq \sum_{n=0}^{n_{\rm cut}}  p_{n|\mu} \left(\frac{1}{N}\sum_{u=1}^N \tilde{Y}^{u,{\rm ref}}_{a,b,n}\right) \leq \frac{1}{p_ap_X^Bp_{\mu}N}\sum_{u=1}^N Q^{u,{\rm ref}}_{a,b,\mu}, \quad \forall \mu
	\end{eqnarray*}
	Finally, by proceeding analogously to the fourth and fifth steps described in \cref{subsec:single-photon-yield}, we find that the gains $Q^{u,{\rm ref}}_{a,b,\mu}$ satisfy 
	\begin{equation}
		g_{\delta}^{\rm L}\left(\frac{1}{N}K_{N,\epsilon}^{\rm L}(M_{a,b,\mu})\right)  \leq \frac{1}{N}\sum_{u=1}^N Q_{a,b,\mu}^{u,{\rm ref}} \leq g_{\delta}^{\rm U}\left(\frac{1}{N}K_{N,\epsilon}^{\rm U}(M_{a,b,\mu})\right),
	\end{equation}
	where $M_{a,b,\mu}$ refers to the number of rounds in which Alice selects the settings $a$ and $\mu$, and Bob observes the successful outcome $b$.
	
	\subsection{Comparison with the decoy-state BB84 protocol}
\begin{figure}[ht]
		\centering
		\includegraphics[width=0.65\textwidth]{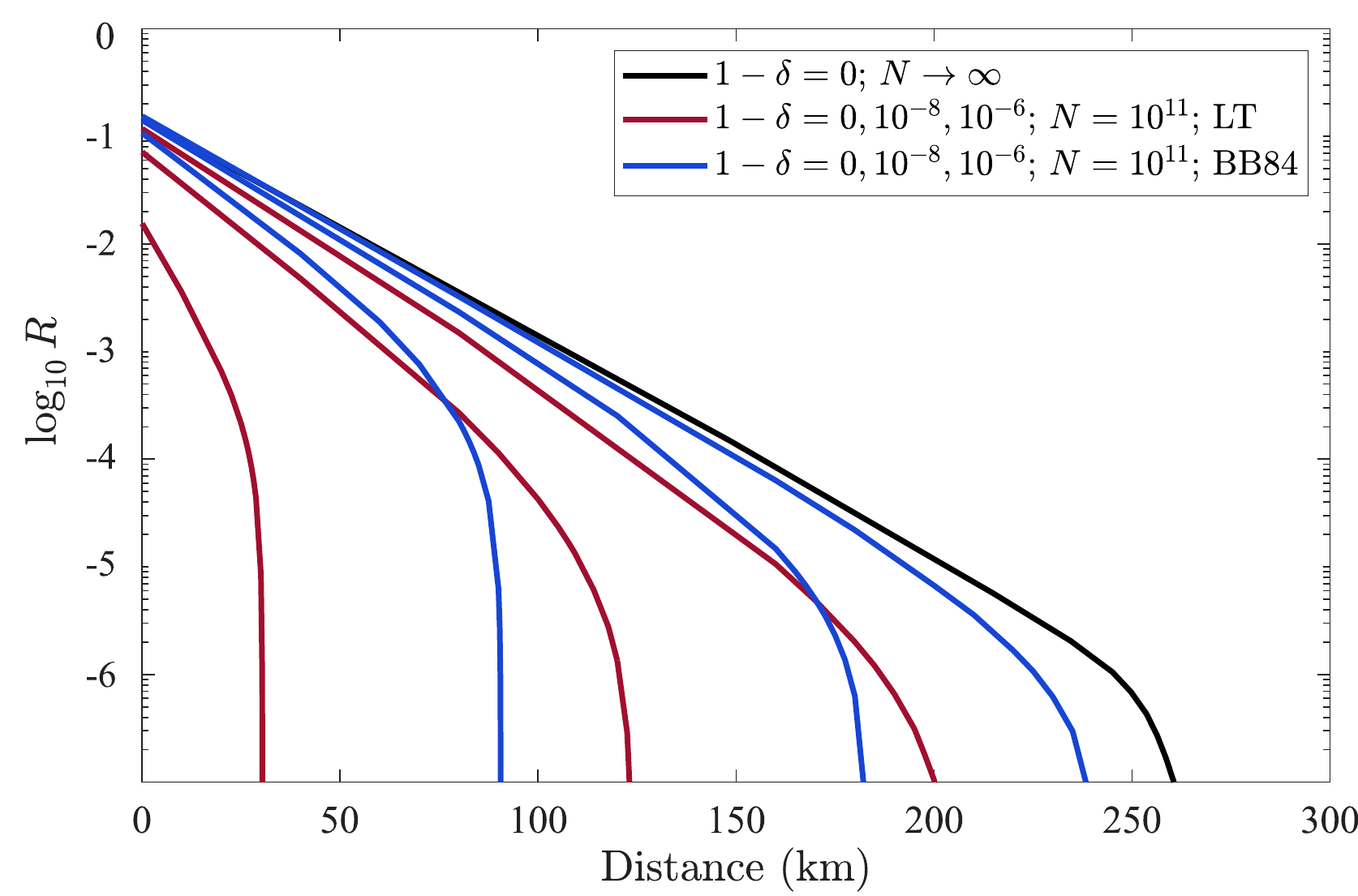}
		\caption{Secret-key rate $R$ of the decoy-state LT protocol (red lines) and the decoy-state BB84 protocol (blue lines) in logarithmic scale for $N=10^{11}$ and for different values of the parameter $\delta$. The figure also includes the asymptotic regime with $\delta=0$ (black line), for which the performance of both schemes coincide. The experimental parameters are those used in \cref{fig:mainFig}.} \label{fig:comparison}
	\end{figure}

	Here we compare the secret-key rate of the  decoy-state LT and BB84 protocols in the presence of information leakage. For the simulations we consider the same experimental and user parameters employed in \cref{fig:mainFig}, with the only exception of $\epsilon$, which in the case of the LT protocol is set to $\epsilon=\varepsilon/32$ since the concentrations bounds must be applied more times than in the BB84 case. The results are shown in~\cref{fig:comparison}, which demonstrates that the BB84 protocol outperforms the LT in all the considered scenarios. This is mainly due to the fact that the phase-error rate estimation of the LT protocol requires to estimate four different yields, requiring to apply the corresponding decoy-state, CS, and concentration bounds more times, while in the BB84 protocol the number of phase errors can be estimated in a more direct way due to the symmetries in the set of transmitted states.

	\section{THA that modifies the intensity of Alice's pulses}\label{appendix:Intensity}
	Here we consider the case in which the photon number statistics of Alice's transmitted pulses may vary with respect to those of the ideal scenario. This might be provoked, for instance, by Eve's injected light, which apart from leaking information about Alice's setting choices through the corresponding back-reflected light, it can also modify the functioning of the laser source increasing the intensity of Alice's pulses~\cite{huang2019laser}, or by passive intensity fluctuations in the transmitted pulses.
	
	Below we shall focus on the case of an active attack, although the analysis is also valid for the case of intensity fluctuations. In particular, we consider that the photon-number statistics of Alice's signal in the $u$-th round no longer satisfy $\tilde{p}_{n|\mu}=p_{n|\mu}$, but instead we have $\tilde{p}_{n|\mu}=p_{n|\tilde{\mu}_u}=e^{-\tilde{\mu}_u}\frac{\tilde{\mu}^n_u}{n!}$, with $\tilde{\mu}_u\in[\mu,\kappa\mu]$, being $\kappa$ a multiplicative factor that depends on Eve's attack. Let us remark, however, that the analysis below could be straightforwardly adapted to any probability distribution. Besides, we consider, as in the main text, that the intensity $\beta_{a,\mu}^2$ of the back-reflected light is upper-bounded by $I_{\max}$ and Eve uses this light to learn information about the settings $a$ and $\mu$. This means, according to~\cref{eq:delta1}, that
	\begin{equation}\label{eq:kappa_delta}
		\begin{split}
			\delta_u & =\sum_{a,\mu}p_{a,\mu}\braket{\tau}{\beta_{a,\mu}e^{i\theta_{a,\mu}}}_E\sum_n \sqrt{p_{n|\mu}p_{n|\tilde{\mu}_u}} \\
			&=\sum_{a,\mu}p_{a,\mu}e^{-\frac{\beta_{a,\mu}^2+\mu+\tilde{\mu}_u-2\sqrt{\mu\tilde{\mu}_u}}{2}} \\
			&\geq \min_ {a,\tilde{\mu}\in[\mu,\kappa\mu]}\left[\sum_{\mu}p_{a,\mu}e^{-\frac{\beta_{a,\mu}^2+\mu+\tilde{\mu}_u-2\sqrt{\mu\tilde{\mu}_u}}{2}}\right]\\
			&\geq\sum_{\mu}p_{\mu}e^{-\frac{I_{\max}+(1+\kappa-2\sqrt{\kappa})\mu}{2}}.
		\end{split}
	\end{equation}
	\begin{figure}[ht]
		\centering
		\includegraphics[width=0.65\textwidth]{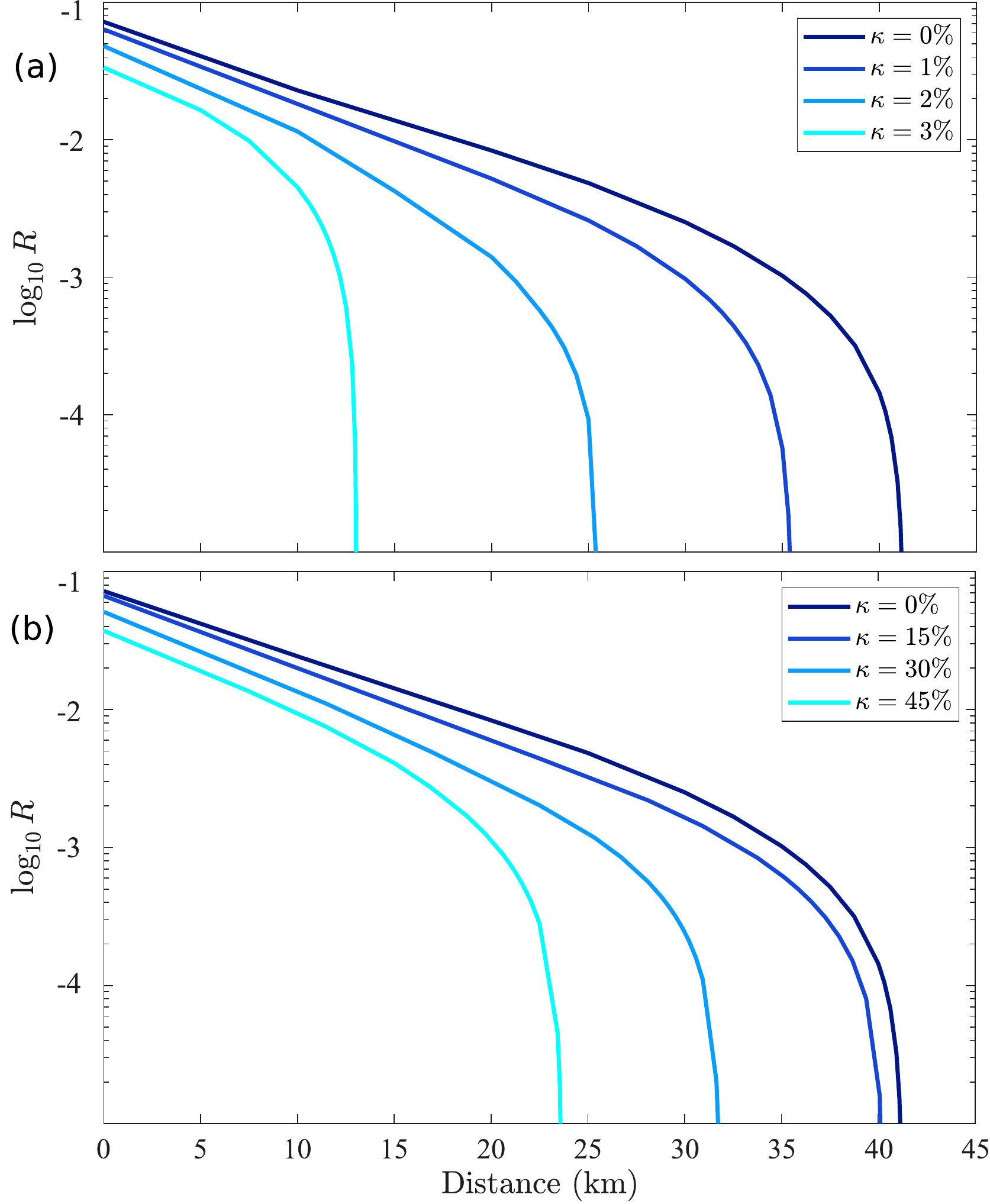}
		\caption{Secret-key rate $R$ of the BB84 protocol in logarithmic scale considering a THA in which Eve injects strong coherent light that, apart from provoking information leakage, it also increases Alice's intensities by a factor $\kappa$, expressed as a percentage. In (a) we assume that the modified intensities could be round dependent (which corresponds to $N_{\rm it}=1$), while in (b) it is assumed that Eve's attack increases the intensity of all the transmitted signals in the same factor $\kappa$ (for which we set, in particular, $N_{\rm it}=16$). We consider that the intensity of Eve's back-reflected light is upper bounded by $I_{\max}=10^{-5}$, and the number of transmitted signals is $N=10^{10}$. The experimental parameters used for the simulation are the same as those used in \cref{fig:mainFig}.} \label{fig:kappa}
	\end{figure}
	
	The results are illustrated in~\cref{fig:kappa}a, which shows how the secret-key rate is significantly affected by the multiplicative factor $\kappa$. In particular, we observe that a small increase in $\kappa$ leads to a quick drop of the secret-key rate. This drop is more notorious if lower values of $I_{\max}$ are selected. This is because the quantities $I_{\max}$ and $(1+\kappa-2\sqrt{\kappa})\mu$ are directly summed in the exponential term of~\cref{eq:kappa_delta}, which means that the increase of $\kappa$ has negligible impact when $\kappa$ changes within a region such that $I_{\max}\gg(1+\kappa-2\sqrt{\kappa})\mu$. Note, however, that one expects $\kappa$ to be small if the isolation at Alice's transmitted is sufficiently high, which is also required to minimize the information leakage.
	
	Significantly better results can be obtained if one assumes that the values of the modified intensities $\tilde{\mu}_u$ do not depend on the particular round, \ie, Eve's attacking strategy is round-independent. In this case, one can consider the worst-case scenario for the secret-key length given that $\tilde{\mu}_u=\tilde{\mu}\in[\mu,\kappa\mu]$. That is, one could take the smallest value of the secret-key length, namely $l_{\rm wcs}$, such that $l_{\rm wcs}\leq\min_{\tilde{\mu}\in[\mu,\kappa\mu]}l_{\tilde{\mu}}$, where $l_{\tilde{\mu}}$ is the secret-key length given in~\cref{eq:lenght}, which now depends on the modified intensities $\tilde{\mu}$ but assumes that these intensities are known precisely. 
	
	Importantly, even though we do not know the exact value of $\min_{\tilde{\mu}\in[\mu,\kappa\mu]}l_{\tilde{\mu}}$, we can obtain a valid $l_{\rm wcs}$ by means of a simple numerical evaluation without compromising the security of the protocol. For this, note that Alice and Bob can always divide the interval $[\mu,\kappa\mu]$ in $N_{\rm it}$ equally-spaced sub-intervals $[\mu_k,\mu_k\kappa_k]$, and take
	\begin{equation}\label{eq:lenght_k}
		l_{\rm wcs}=\min_{k=0,\dots,N_{\rm it}-1}l_{[\mu_k,\mu_k\kappa_k]},
	\end{equation}
	where $l_{[\mu_k,\mu_k\kappa_k]}$ is the secret-key length obtained by considering the intensity settings $\mu_k=\mu+k\Delta\mu$, with $\Delta\mu=\frac{\mu(\kappa-1)}{N_{\rm it}}$, and a multiplicative factor $\kappa_k=\frac{\mu+(k+1)\Delta\mu}{\mu+k\Delta\mu}=\frac{N_{\rm it}+(k+1)(\kappa-1)}{N_{\rm it}+k(\kappa-1)}$. That is, for each $k$, the imperfections due to Eve's attack, are incorporated through~\cref{eq:kappa_delta}, but substituting $\mu\to\mu_k$ and $\kappa\to\kappa_k$ in that equation. Note that, in doing so, one gets as close as desired to the perfect minimization $\min_{\mu\in[\mu,\kappa\mu]}l_{\mu}$ by increasing $N_{\rm it}$ without compromising in any case the security of the protocol. The results for the particular case of $N_{\rm it}=16$ and for different values of $\kappa$ (expressed as percentage) are shown in~\cref{fig:kappa}b. As expected, the secret-key rate is much less sensitive to Eve's attack than in the previous scenario shown in~\cref{fig:kappa}a.

	\section*{References}
	\bibliographystyle{iopart-num}
	\bibliography{refs}

\providecommand{\newblock}{}
\begin{thebibliography}{10}
\expandafter\ifx\csname url\endcsname\relax
  \def\url#1{{\tt #1}}\fi
\expandafter\ifx\csname urlprefix\endcsname\relax\def\urlprefix{URL }\fi
\providecommand{\eprint}[2][]{\url{#2}}

\bibitem{scarani}
Scarani V, Bechmann-Pasquinucci H, Cerf N~J, Du{\v s}ek M, L{\"u}tkenhaus N and
  Peev M 2009 {\em Rev. Mod. Phys.\/} {\bf 81} 1301

\bibitem{lo}
Lo{\;}H-K, Curty M and Tamaki K 2014 {\em Nat. Photonics\/} {\bf 8} 595--604

\bibitem{portmann2021security}
Portmann C and Renner R 2021 {\em preprint arXiv:2102.00021\/}

\bibitem{bennett}
Bennett C~H and Brassard G 1984 {\em Proc. IEEE Int. Conf. on Computers,
  Systems and Signal Processing (Bangalore, India)\/} pp 175--179

\bibitem{vernam1926}
Vernam G~S 1926 {\em J. Am. Inst. Electr. Eng.\/} {\bf 45} 295--301

\bibitem{mayers2004self}
Mayers D and Yao A 2004 {\em Quantum Information \& Computation\/} {\bf 4}
  273--286

\bibitem{acin2007device}
Ac{\'\i}n A, Brunner N, Gisin N, Massar S, Pironio S and Scarani V 2007 {\em
  Phys. Rev. Lett.\/} {\bf 98} 230501

\bibitem{vazirani2014fully}
Vazirani U and Vidick T 2014 {\em Phys. Rev. Lett.\/} {\bf 113}(14) 140501

\bibitem{arnon2018practical}
Arnon-Friedman R, Dupuis F, Fawzi O, Renner R and Vidick T 2018 {\em Nat.
  Commun.\/} {\bf 9} 1--11

\bibitem{gisin2}
Gisin N, Fasel S, Kraus B, Zbinden H and Ribordy G 2006 {\em Phys. Rev. A\/}
  {\bf 73} 022320

\bibitem{vakhitov}
Vakhitov A, Makarov V and Hjelme D~R 2001 {\em J. Mod. Opt.\/} {\bf 48} 2023

\bibitem{huang2019laser}
Huang A, Navarrete {\'A}, Sun S~H, Chaiwongkhot P, Curty M and Makarov V 2019
  {\em Phys. Rev. App.\/} {\bf 12} 064043

\bibitem{lucamarini2015Practical}
Lucamarini M, Choi I, Ward M~B, Dynes J~F, Yuan Z~L and Shields A~J 2015 {\em
  Phys. Rev. X\/} {\bf 5} 031030

\bibitem{tamaki2016decoy}
Tamaki K, Curty M and Lucamarini M 2016 {\em New J. Phys.\/} {\bf 18} 065008

\bibitem{wang2018finite}
Wang W, Tamaki K and Curty M 2018 {\em New J. Phys.\/} {\bf 20} 083027

\bibitem{pereira2019quantum}
Pereira M, Curty M and Tamaki K 2019 {\em npj Quantum Inf.\/} {\bf 5} 62

\bibitem{hwang2003quantum}
Hwang{\;}W-Y 2003 {\em Phys. Rev. Lett.\/} {\bf 91} 057901

\bibitem{lo2005decoy}
Lo{\;}H-K, Ma X and Chen K 2005 {\em Phys. Rev. Lett.\/} {\bf 94} 230504

\bibitem{wang2005beating}
Wang{\;}X-B 2005 {\em Phys. Rev. Lett.\/} {\bf 94} 230503

\bibitem{wang2021measurement}
Wang W, Tamaki K and Curty M 2021 {\em Scientific Reports\/} {\bf 11} 1--11

\bibitem{tamaki2014loss}
Tamaki K, Curty M, Kato G, Lo{\;}H-K and Azuma K 2014 {\em Phys. Rev. A\/} {\bf
  90} 052314

\bibitem{kato}
Kato G 2020 {\em preprint arXiv:2002.04357\/}

\bibitem{pereira2020quantum}
Pereira M, Kato G, Mizutani A, Curty M and Tamaki K 2020 {\em Science
  Advances\/} {\bf 6} eaaz4487

\bibitem{navarrete2021practical}
Navarrete {\'A}, Pereira M, Curty M and Tamaki K 2021 {\em Phys. Rev. App.\/}
  {\bf 15} 034072

\bibitem{zhou2016making}
Zhou Y~H, Yu Z~W and Wang{\;}X-B 2016 {\em Phys. Rev. A\/} {\bf 93} 042324

\bibitem{lim2014concise}
Lim C~C~W, Curty M, Walenta N, Xu F and Zbinden H 2014 {\em Phys. Rev. A\/}
  {\bf 89} 022307

\bibitem{zhang2017improved}
Zhang Z, Zhao Q, Razavi M and Ma X 2017 {\em Phys. Rev. A\/} {\bf 95} 012333

\bibitem{zapatero2021security}
Zapatero V, Navarrete {\'{A}}, Tamaki K and Curty M 2021 {\em {Quantum}\/} {\bf
  5} 602

\bibitem{jensen1906fonctions}
Jensen J~L~W~V 1906 {\em Acta Mathematica\/} {\bf 30} 175--193

\bibitem{yin2016measurement}
Yin H~L {\em et~al.\/} 2016 {\em Phys. Rev. Lett.\/} {\bf 117}(19) 190501

\bibitem{curras2021tight}
Curr{\'a}s-Lorenzo G, Navarrete {\'A}, Azuma K, Kato G, Curty M and Razavi M
  2021 {\em npj Quantum Inf.\/} {\bf 7} 1--9

\bibitem{curras2021finite}
Curr\'as-Lorenzo G, Navarrete {\'A}, Pereira M and Tamaki K 2021 {\em Phys.
  Rev. A\/} {\bf 104}(1) 012406

\bibitem{tomamichel2012tight}
Tomamichel M, Lim C~C~W, Gisin N and Renner R 2012 {\em Nat. Commun.\/} {\bf 3}
  1--6

\bibitem{tomamichel2011leftover}
Tomamichel M, Schaffner C, Smith A and Renner R 2011 {\em IEEE Trans. Inform.
  Theory\/} {\bf 57} 5524--5535

\bibitem{vitanov2013chain}
Vitanov A, Dupuis F, Tomamichel M and Renner R 2013 {\em IEEE Trans. Inform.
  Theory\/} {\bf 59} 2603--2612

\bibitem{tomamichel2011uncertainty}
Tomamichel M and Renner R 2011 {\em Phys. Rev. Lett.\/} {\bf 106} 110506

\end{thebibliography}

\end{document}